# Modeling meso-scale energy localization in shocked HMX, Part I: machine-learned surrogate model for effect of loading and void size

A. Nassar, N. K. Rai, O. Sen, and H.S. Udaykumar[*]

*Mechanical and Industrial Engineering, The University of Iowa, Iowa City, Iowa 52242*

Abstract

This work presents the procedure for constructing a machine learned surrogate model for hotspot ignition and growth rates in pressed HMX materials. A Bayesian Kriging algorithm is used to assimilate input data obtained from high-resolution meso-scale simulations. The surrogates are built by generating a sparse set of training data using reactive meso-scale simulations of void collapse by varying loading conditions and void sizes. Insights into the physics of void collapse and ignition and growth of hotspots are obtained. The criticality envelope for hotspots is obtained as the function $\Sigma_{cr} = f(P_s, D_{void})$ where $P_s$ is the imposed shock pressure and $D_{void}$ is the void size. Criticality of hotspots is classified into the plastic collapse and hydrodynamic jetting regimes. The information obtained from the surrogate models for hotspot ignition and growth rates and the criticality envelope can be utilized in meso-informed Ignition and Growth (MES-IG) models to perform multi-scale simulations of pressed HMX materials.

## 1. INTRODUCTION

Heterogeneous energetic materials (HEs) have complex meso-structures replete with defects such as cracks, voids and interfaces [1]. Sensitivity and initiation of shock-loaded HEs hinge on the dynamics at the meso-scale; shock-focusing and localization of energy at defects lead to hotspots [2, 3]. Reactions initiate at the hotspots and spread into the unreacted material. If the reaction fronts propagate sufficiently rapidly, they can catch up to the shock; coupling between the reaction and shock fronts can then lead to a detonation wave. The shock-to-detonation transition (SDT) therefore hinges on the strength of the shock loading (measured by the shock pressure $P_s$ and pulse duration $\tau_s$) and features of the meso-scale morphology, chiefly the porosity $\phi$ and the average void size $D_{void}$. Void shape and orientation can also have strong effects on initiation sensitivity, as shown in recent work [4-6]. This work focuses on the effects of shock loading ($P_s$, $\tau_s$) and void size $D_{void}$ on the energy localization at the meso-scale. A companion paper [7] addresses the effects of void volume fraction $\phi$ and the shape/orientation of voids on hotspot intensity and HE initiation sensitivity.

Sen et al. [8] simulated the shock-to-detonation transition in pressed HMX materials using a meso-informed macro-scale shock hydrodynamics framework. The macro-scale system of equations was closed using an ignition-and-growth model for energy localization, in a framework called MES-IG. In MES-IG [8], unlike in phenomenological ignition and growth models [9], the reaction progress variable was obtained from a surrogate model. The particular terms in MES-IG [8] that capture the meso-scale hotspot mechanisms are the macro-scale reaction progress rates $\dot{\lambda}_{ignition} = \phi \dot{F}_{ignition}$, and $\dot{\lambda}_{growth} = \phi \dot{F}_{growth}$ where $\lambda$ is the conventional [9] reacted mass fraction at the macro-scale. $F$ is the reacted mass fraction at the meso-scale, i.e., for a single, isolated void and $\phi$ is the void volume fraction. Traditional ignition and growth modeling uses the rates $\dot{\lambda}_{ignition}$ and $\dot{\lambda}_{growth}$ to model the heat release in a macro-scopic control volume. However, the ignition and growth model relies on empirical fits [9] to obtain these rates. The current work injects information regarding meso-scale hotspot dynamics into the ignition and growth model. In the MES-IG framework, surrogate models for $\dot{F}_{\text{ignition}}$ and $\dot{F}_{\text{growth}}$ are obtained by training the Bayesian Kriging estimator [10-14] using high-resolution reactive meso-scale simulations of void collapse. In this paper, we describe the procedures used for obtaining the training or input data for constructing the surrogates: $\dot{F}_{\text{ignition}} = \dot{F}_{\text{ignition}}^{\text{scv}}(P_s, \tau_s, D_{\text{void}})$ and $\dot{F}_{\text{growth}} = \dot{F}_{\text{growth}}^{\text{scv}}(P_s, \tau_s, D_{\text{void}})$, where the subscript *scv* indicates that meso-scale computational experiments are performed on a single cylindrical void. This surrogate model captures the effect of loading and void size on the energy localization at the meso-scale, which is the unresolved or subgrid scale from the macro-scale perspective. One disadvantage of machine learning from ensembles of meso-scale simulations is that the resulting surrogate or closure model, while useful, can lack interpretability, i.e., the interesting physics at the meso-scale can be lost. In this work, we connect the physics of meso-scale hot spot formation to the hypersurfaces generated for $\dot{F}_{\text{ignition}}$ and $\dot{F}_{\text{growth}}$.

---

[*] Corresponding author; email : ush@engineering.uiowa.edu

## 1.1. Particle and void size effects on sensitivity of energetic materials

The primary goal of this work is to infuse macro-scale HE simulations with information from the meso-scale, specifically with regard to the effect of void size on energy localization. A variety of studies have examined the importance of particle and defect (e.g. pore/void) size on the sensitivity of HEs. In an early work by Price [15], synthesizing the experimental observations available at the time, the effect of particle size on sensitivity was examined for a range of experiments on different materials. Price [15] noted that wedge test experiments on TNT (Campbell et al., [16]) indicated that fine TNT is more sensitive. In contrast, gap test experiments by Dinegar et al. [17] on PETN indicated that coarse particles enhanced sensitivity. Other gap test experiments following Dinegar [17] also showed coarse HEs to be more sensitive. For example, Scott [18] showed that, for RDX and PETN, coarser particles ignite more easily, i.e., at lower initial shock pressure. But once ignited, fine particles showed more rapid buildup to detonation than coarse particles. This distinction between coarse and fine particle HEs was observed in later work as well, for example by Howe et al. [19]. Thus, experiments show a cross-over behavior exhibited by HEs comprising coarse and fine particles, as illustrated by Price [15] and shown in Figure 1. At loading pressures $P_s < P_r$ the ease of ignition determines the response and the coarse material appears to be more sensitive. At $P_s > P_r$ ignition will occur for the coarse and fine materials, but the rate of reaction predominates and the fine material appears to be more sensitive than the coarse.

Price [15] resolves the conflict between the gap and wedge test data by noting that the two types of tests could be on either side of $P_r$. Most gap tests on porous HEs used low amplitude, long duration shocks while most foil flyer impact tests used high amplitude, short duration shocks. Therefore, coarse particle HEs are more sensitive in gap tests, while fine particle HEs are more sensitive in flyer impact tests. This difference in the response to loading rates was demonstrated in experiments. Schwarz [20] showed that the Walker-Wasley relation for criticality is approximately satisfied for short-pulse high intensity loading; they obtained the criticality relationship $P_s^{2.4} \tau_s = $ constant for such cases and asserted that fines are more sensitive. Fine particles have also been shown to be more sensitive than coarse in pressed HMX by more recent experiments of Welle and Molek [21]. Honodel et al. (1981) [22] did both flyer and gap tests on TATB to show that for thin flyer, high pressure cases fines are more sensitive, while for thick flyer and gap tests coarse materials are more sensitive. While for higher pressures they showed that $P_s^2 \tau_s = $ constant roughly holds, for lower pressure loadings this relationship was not valid. Therefore, the relative sensitivity of fine/coarse particle HEs, corresponding to HEs with small/large voids respectively, depends in a complex way on the loading regime. A meso informed macro-model, therefore, should correctly reflect these sensitivity characteristics of coarse/fine particle HEs.

The main mechanistic aspects of particle (and thereby void) size effects are fairly well understood. For instance, Honodel et al. [22] explained their observations on particle size effects in the following way. Larger voids lead to larger hotspots upon collapse. Larger hotspots survive longer than smaller hotspots.

Also, the energy localized in a hotspot is proportional to the loading intensity. Hence, at lower pressures the coarse material will ignite more easily because void collapse will result in low temperature hotspots; the larger hotspots will localize more energy and will survive long enough to trigger reactions. Therefore, at low pressures coarser HEs, which contain larger particles and larger voids, will be more sensitive. On the other hand, at high pressures, collapse of voids will result in more intense hotspots and the total number of hotspots will predominate over the size of hotspots in determining the time of reaction. At higher pressures, for a given void fraction, fine HE will contain more voids and will therefore appear to be more sensitive than coarse HE. This cross-over of sensitivity between larger and smaller void sizes was demonstrated in the run-to-detonation simulations and the Walker-Wasley curves obtained from the meso-informed ignition and growth (MES-IG) model in Sen et al. [8]. In this work, further insights into the effects of void size at various shock loads will be obtained by scanning a large swathe of the parameter space $(P_s, \tau_s, D_{\text{void}})$.

Using a Carroll-Holt (spherical collapse) model [23], Frey [24] investigated void sensitivity-size relationships. He states that the maximum energy available to be dissipated (for spherical collapse) is the product of the cavity size and the applied pressure. Frey also states the following: "in a purely hydrodynamic model, with plastic work and gas phase heating ignored, the temperature which can be achieved at a cavity is independent of cavity size if the shock rise time is infinitesimal. Therefore, the size of the hotspot will increase with cavity size but the temperature does not. However, if plastic work is considered, cavity size has a strong effect on the results." However, Frey's [24] conjectures may be influenced by assumptions of spherical collapse in the Carroll-Holt model [23]. In the shock-loaded collapse of a void, there is an intrinsic directionality due to shock propagation. The Carroll-Holt spherical compression model then only approximates the situation for plastic deformation of the void and the "void closure" mode of collapse. The hydrodynamic "jetting" mode is not represented in a spherical collapse model. Thus, by removing the geometric constraints of the Carroll-Holt model [23, 25], the characteristics of hotspots (e.g. hotspot size $D_{\text{hs}}$ and temperature $T_{\text{hs}}$) and their dependency on the loading rates can be examined over a wider parameter space, which runs across the plasticity and hydrodynamics dominated regimes.

Despite the above general physical explanations of the effect of void size on sensitivity, there are some aspects of void size effects that remain puzzling. For example, Honodel at al. [22] and Moulard et al. [26] have shown that there is a size-dependent sensitivity reversal in the high pressure loading regime. Initially, when particle size is decreased sensitivity increases. Honodel et al. [22] and Khasainov et al. [27] attribute this to increased surface area for finer voids. However, when the particle size is deceased below a certain value the sensitivity again decreases. This is argued to be because, for extremely small voids, the hot spot is too small to cause ignition for a given load. In the present work, we show that there is indeed a reversal in sensitivity with respect to void size; there is an optimal void size range of about 1-10 μm [28] where the most sensitive voids can be found. Very small voids are quenched by diffusion, while very large voids (size of 10s of μm) are not sufficiently abundant in a meso-structure of a given void fraction and therefore not as effective as voids in the 1-10 μm range. In the range of interest as considered in this paper i.e., $D_{\text{void}} = $ 1μm-10μm, we find that there is a non-monotonic dependency of sensitivity on void size. The physics underlying this non-monotonicity is elucidated in this paper.

### 1.2. The critical hotspot size

Discussion of the most effective hotspot size is underpinned by the concept of a critical hotspot, developed by Tarver, Chidester and Nichols [29]. Tarver et al. [29] assumed a spherical hotspot of a certain size (and uniform temperature) in an otherwise unreacted material (HMX). By solving a reaction-diffusion problem Tarver et al. [29] asked: for what size and intensity of hotspot does the surrounding HMX completely react? If there is insufficient energy in the hotspot (i.e., the hotspot is too small or of low intensity, i.e., temperature) diffusion may transport heat away from the hotspot before complete consumption of the HMX can ensue. Tarver and coworkers [29] determined the boundary between the go-no-go regions, corresponding to "critical" hotspots. Large hotspots (millimeter sized ones that may arise in impact scenarios) may take much longer to reach criticality. Smaller hotspots (μm-sized) that arise in shocked materials take much shorter times to reach explosion (order of micro-seconds or even sub-microseconds). Note that the smaller the critical hotspots, the higher will be their temperature.

While the Tarver hotspot curve [29] indicates what will happen to a hotspot (i.e., whether it will go critical or not) it does not indicate how that hotspot may have arisen in the first place. It is important to know: what size void must collapse under what loading condition in order to form a hotspot of critical size (and temperature)? This is an issue that is addressed in this paper by examining the regions in the parameter space – defined by shock loading $(P_s, \tau_s)$ and void size $D_{\text{void}}$ – where void collapse is sub- and super-critical. This allows for delineation of the criticality hypersurface $\Sigma_{\text{cr}} = \Sigma\left(P_s, \tau_s, D_{\text{void}}\right)$. This criticality criterion has the advantage of being defined in terms of operating conditions of a HE sensitivity experiment, rather than in terms of hotspot size and temperature. Such a criticality hypersurface can be useful for design of HE sensitivity.

This paper develops models for the ignition and growth rates of reactive hotspots at the mesoscale; reactions are initiated at sites of void collapse. Simulations of reactive single cylindrical void (*scv*) collapse are used to extract the information needed to construct surrogate models in the form $\dot{F}_{\text{ignition}} = \dot{F}^{\text{scv}}_{\text{ignition}}(P_s, \tau_s, D_{\text{void}})$ and $\dot{F}_{\text{growth}} = \dot{F}^{\text{scv}}_{\text{growth}}(P_s, \tau_s, D_{\text{void}})$. The inputs to a Bayesian Kriging surrogate model [10, 11], are obtained by performing an ensemble of meso-scale void collapse simulations spanning the parameter space $(P_s, \tau_s, D_{\text{void}})$. In the following section, the equations solved to perform meso-scale single void collapse calculations are briefly described (Section 2). Detailed implementational and other modeling issues for such simulations are presented in Rai et al. [30, 31]. Section 2 describes the techniques used to extract input data from the simulations for the Bayesian learning algorithm. Section 3 presents the results obtained and the physical insights provided by the ensemble of meso-scale simulations. The comprehensive span of the parameter space, involving ~ 600 high-resolution void-collapse simulations, lead not only to the construction of surrogate models, but also to insights into the complex issues related to void size and loading effects mentioned in the introduction section of this paper. Finally, Section 4 provides the conclusions reached by the present work.

## 2. METHODS

### 2.1. Governing Equations and Solution Techniques

At the mesoscale, the governing equations for the conservation of mass, momentum and energy are given by the following set of hyperbolic equations:

$$\frac{\partial \rho}{\partial t} + \frac{\partial (\rho u_i)}{\partial x_i} = 0 \tag{1}$$

$$\frac{\partial (\rho u_i)}{\partial t} + \frac{\partial (\rho u_i u_j - \sigma_{ij})}{\partial x_j} = 0 \tag{2}$$

and

$$\frac{\partial (\rho E)}{\partial t} + \frac{\partial (\rho E u_j - \sigma_{ij} u_i)}{\partial x_j} = 0 \tag{3}$$

where $\rho$, and $u_i$, respectively denote the density, and the velocity components, $E = e + \frac{1}{2} u_i u_i$ is the specific total energy, and $e$ is the specific internal energy of the mixture. The Cauchy stress tensor, $\sigma_{ij}$, is decomposed into volumetric and deviatoric components, i.e.,

$$\sigma_{ij} = S_{ij} - p\delta_{ij} \qquad (4)$$

The deviatoric stress tensor, $S_{ij}$, is evolved using the following evolution equation:

$$\frac{\partial(\rho S_{ij})}{\partial t} + \frac{\partial(\rho S_{ij} u_k)}{\partial x_k} + \frac{2}{3}\rho G D_{kk}\delta_{ij} - 2\rho G D_{ij} = 0 \qquad (5)$$

where $D_{ij}$ is the strain rate tensor, and $G$ is the shear modulus of material. First, the deviatoric stresses are evolved by the elastic update above and then mapped back to the yield surface using a radial return algorithm [32]. The yield surface is given by the function $f = S_\text{e} - \sigma_\text{y}$, where $S_\text{e} = \sqrt{\frac{3}{2}(S_{ij}S_{ij})}$. The yield strength, $\sigma_\text{y}$ is taken to be a constant and set to 260 MPa for HMX [33], i.e., hardening, and visco-plastic effects are neglected in the mesoscale computational models. A detailed description of the governing equations and the radial return algorithm is provided in previous works [34, 35].

Void collapse can lead to the melting of HMX; therefore, thermal softening of HMX is modeled using the Kraut-Kennedy relation, $T_\text{m} = T_{m0}\left(1 + a\frac{\Delta V}{V_0}\right)$, with model parameters provided in the work of Menikoff et al. [33]. Once the temperature exceeds the melting point of HMX, the deviatoric strength terms are set to zero. Furthermore, the specific heat of HMX is known to change significantly with temperature. The variation of specific heat is modeled as a function of temperature as suggested in [36].

The pressure at the mesoscale is obtained from a Birch-Murnaghan equation of state [33, 36], which can be written in the general Mie-Gruneisen form as:

$$p(\rho, e) = p_\text{k}(\rho) + \rho\Gamma_\text{s}[e - e_\text{k}(\rho)] \qquad (6)$$

where,

$$p_\text{k}(\rho) = \frac{3}{2}K_{T0}\left[\left(\frac{\rho}{\rho_0}\right)^{\frac{7}{3}} - \left(\frac{\rho}{\rho_0}\right)^{\frac{5}{3}}\right]\left[1 + \frac{3}{4}(K'_{T0} - 4)\left[\left(\frac{\rho}{\rho_0}\right)^{\frac{2}{3}} - 1\right]\right] \qquad (7a)$$

$$e_\text{k}(\rho) = e_0 - \int_{\frac{1}{\rho_0}}^{\frac{1}{\rho}} p_\text{k}(\rho)\, d\left(\frac{1}{\rho}\right) \qquad (7b)$$

The governing equations of mass, momentum and energy along with evolution of deviatoric stresses are spatially discretized using a 3rd-order essentially non-oscillatory scheme [37] and numerically integrated using 3rd-order Runge-Kutta time stepping. The interfaces in the current framework are modeled using the narrow-band levelset approach [38]. The levelset approach allows for tracking the interfaces in a sharp manner and can efficiently handle the large deformation of materials. The interfacial conditions are implicitly satisfied using a modified ghost fluid method [37]. The complete description of the numerical framework can be found in previous work [34, 35, 39, 40].

## 2.2. Reactive Modeling of HMX

Chemical decomposition of HMX is modeled in the Tarver 3-equation model [29] via four different species. The three reactions are:

<u>Reaction 1:</u>  HMX ($C_4H_8N_8O_8$) → fragments ($CH_2NNO_2$)  (8)

Reaction 2:  fragments ($CH_2NNO_2$) → intermediate gases ($CH_2O, N_2O, HCN, HNO_2$)   (9)

and

Reaction 3:  $2 \times$ intermediate gases ($CH_2O, N_2O, HCN, HNO_2$)
→ final gases ($N_2, H_2O, CO_2, CO$)   (10)

The solid HMX (species 1, with mass fraction $Y_1$) at high temperature decomposes to fragments (species 2, with mass fraction $Y_2$). The fragments further decompose to intermediate gases (species 3 with mass fraction $Y_3$) and finally the intermediate gases undergo strong exothermic reactions to form gaseous products (species 4, $Y_4$), which leads to an increase in temperature.

The chemical species are evolved in time by solving the species conservation equation:

$$\frac{\partial \rho Y_i}{\partial t} + \text{div}(\rho \vec{V} Y_i) = \dot{Y}_i \qquad (11)$$

where $Y_i$ is the mass fraction of the $i^{th}$ species and $\dot{Y}_i$ is the production rate source term for the $i^{th}$ species. The numerical stiffness in solving the reactive set of equations is circumvented by using a Strang operator-splitting approach [41], where first the advection of species is performed using the flow time step to obtain predicted species values:

$$\frac{\partial \rho Y_i^*}{\partial t} + \text{div}(\rho \vec{V}^n Y_i^*) = 0 \qquad (12)$$

In a second step, the evolution of the species mass fraction due to chemical reactions is calculated:

$$\frac{dY_i^{n+1}}{dt} = \dot{Y}_i^* \qquad (13)$$

The species evolution in Eqn. (13) is advanced in time using a $5^{th}$-order Runge-Kutta Fehlberg [42] method, which uses an internal adaptive time-stepping scheme to deal with the stiffness of the chemical kinetics. Further details on the reactive mechanics calculations are available in previous work [30].

The heat released due to the decomposition of solid HMX into the gaseous products is modeled via the 3-equation model of Tarver et al. [29] as explained using Eqn. (8-10). The change in temperature because of the chemical decomposition of HMX is calculated by solving the evolution equation,

$$\rho C_p \dot{T} = \dot{Q}_R + k \nabla^2 T \qquad (14)$$

where $\rho$ is the density of HMX, $C_p$ is the specific heat of HMX at constant pressure, $T$ is the temperature, $k$ is the thermal conductivity of HMX and $\dot{Q}_R$ is the total heat release rate of the chemical reaction. The values of $C_p$, $k$ and $\dot{Q}_R$ are obtained from the work of Tarver et al. [29]. Further details on the calculation of reactive void collapse using the Tarver 3-equation chemistry model has been described in detail in previous works [30, 31].

## 2.3. The Training Algorithm for constructing the Surrogate Model

In the present work, high-resolution mesoscale simulations are used to construct surrogate models for the hot-spot induced ignition and growth rates, $\dot{F}_{ignition} = \dot{F}^{scv}_{ignition}(P_s, \tau_s, D_{void})$ and $\dot{F}_{growth} = \dot{F}^{scv}_{growth}(P_s, \tau_s, D_{void})$. A surrogate model approximates an unknown function $f(x)$, which is known only at a sparse set of discrete and distinct points $x_j$ ($j$=1,…$N$) [11, 43]. The data set which is used for constructing the approximation is known as the input or training data, and is obtained from mesoscale numerical experiments in the present work. The inputs are used to create a fit/hypersurface for the unknown function in the parameter space. The interpolation/fitting function is typically a universal approximator [44], with a prescribed model complexity. The process of estimating the model parameters in the universal approximator is referred to as the training or surrogate construction process.

To train the multi-dimensional surrogate models, $\dot{F}^{scv}_{ignition}$ and $\dot{F}^{scv}_{growth}$, a Modified Bayesian Kriging (MBKG) method is used [10]. MBKG surrogate construction [10, 11] assumes the inputs come from a stationary Gaussian random process, with a mean value of $\boldsymbol{P\lambda} + \boldsymbol{Z}$ and variance $\sigma^2\beta$, i.e.,

$$\tilde{f}(x_0) \sim \text{MVN}(\boldsymbol{P\lambda} + \boldsymbol{Z}, \sigma^2\beta \boldsymbol{I}) \quad (15)$$

where MVN refers to a multivariate normal distribution, $\boldsymbol{P\lambda}$ represents the mean structure and $\boldsymbol{Z}$ is modeled as a Gaussian random process with zero mean and covariance $E[Z(\boldsymbol{x}_j)Z(\boldsymbol{x}_q)] = \sigma^2 \boldsymbol{R}$. The matrix, $\boldsymbol{R}$ is a spatial correlation of the input points, $\boldsymbol{x}_j$ and $\boldsymbol{x}_q$, and is defined as follows:

$$\boldsymbol{R} = R_{jq} = R(\theta, \boldsymbol{x}_j, \boldsymbol{x}_q) = \prod_{k=1}^{n} \gamma_k(\theta_k, d_k) \quad (16)$$

where $\theta$ is a shape parameter, $d_k = (x_{kj} - x_{kq})$, $k = 1,2,…,n$, $n$ being the dimension of the vector $\boldsymbol{x}$. Commonly used models of the correlation functions are listed in [68]. The unknown parameters in the MBKG model are $\boldsymbol{\lambda}$, $\sigma^2$, $\theta$ and $\beta$ and are estimated using Gibbs sampling algorithm and a Markov Chain Monte Carlo Method (MCMC) in a Bayesian learning framework. The priors and the likelihood functions, as well as the convergence criteria and the training method are described in detail in previous work [11].

## 2.4. The Quantities of Interest (QoIs)

The MBKG method [11] is used to supply closures to the macro-scale MES-IG model [8]. The first step in building the MES-IG model is to obtain the training data for the surrogate models:

$$\dot{F}_{ignition} = \dot{F}^{scv}_{ignition}(P_s, \tau_s, D_{void}) * f^{v-v}_{ignition}(\phi) * f^{shape}_{ignition}(AR, \theta) \quad (17a)$$

and

$$\dot{F}_{growth} = \dot{F}^{scv}_{growth}(P_s, \tau_s, D_{void}) * f^{v-v}_{growth}(\phi) * f^{shape}_{growth}(AR, \theta) \quad (17b)$$

where the functions $\dot{F}^{scv}_{ignition}$ and $\dot{F}^{scv}_{growth}$ consider the effect of loading and void size for a single cylindrical void, $f^{v-v}_{ignition}(\phi)$ or $f^{v-v}_{growth}(\phi)$ and $f^{shape}_{ignition}(AR, \theta)$ or $f^{shape}_{growth}(AR, \theta)$ consider the void-void interaction and void shape effects respectively and the subscripts denote the ignition and the growth phases. The latter two functions on the right hand side of Eqns. (17) are obtained in Part II [7]. In this paper, two QoIs are extracted from the data derived from the meso-scale computations. They are the ignition and growth rates, $\dot{F}^{scv}_{ignition}$ and $\dot{F}^{scv}_{growth}$ for different combinations of $P_s, \tau_s,$ and $D_{void}$. The procedure for obtaining $\dot{F}^{scv}_{ignition}$ and $\dot{F}^{scv}_{growth}$ is explained next.

2.5.1. Numerical setup and approach to obtain the training data for the QoIs

The numerical setup used to conduct the mesoscale reactive simulations is shown in Figure 3. The simulations are performed at selected points in a three-dimensional parameter space defined by the shock pressure ($P_s$), shock pulse duration ($\tau_s$), and void diameter ($D_{void}$). To unify the assimilation of data for the range of pressures and diameters, the data is presented by defining a non-dimensional shock pulse duration ($\tau^*$) that is a function of $P_s$ and $D_{void}$:

$$\tau_s^* = \frac{\tau_s}{\frac{D_{void}}{u_s}} \tag{18}$$

where the shock speed ($u_s$) is calculated from:

$$u_s = c_o + s\, u_p \tag{19}$$

where $u_s$ is directly related to shock pressure through the material Hugoniot [33], $u_p$ is the particle velocity in m/s, $c_o$ is the speed of sound in HMX (2740 m/s) and s is a material constant (1.49 for HMX). In each case, the variable $\tau_s^*$ is the normalized shock passage time for a given void diameter. The definition of $\tau_s^*$ allows the surrogate to cover the relevant range of the shock pulse duration for different void sizes and shock velocities.

Figure 3 shows typical results of a case with input parameters $P_s$ = 22.1 GPa, $\tau_s^*$ = 1.5, and $D_{void}$ = 30 µm. The results show the time variation of the void area $A_{void}$, average hotspot temperature $T_{hs}$, hotspot area $A_{hs}$, and product mass fraction $F$ in Figures 4(a-d) respectively. The hotspot is defined as the region where the local temperature is greater than the bulk temperature that results from shock heating. The average hotspot temperature, $T_{hs}$ is the Favre-average temperature in the hotspot region:

$$T_{hs} = \frac{\int_{A_{hs}} \rho T \mathrm{d}A}{\int_{A_{hs}} \rho \mathrm{d}A} \tag{20}$$

where $A_{hs}$ is the hotspot area, $\rho$ is the density and $T$ is the local temperature in the hotspot region.

Figure 4(a) shows that the initial value of $A_{void}$ is approximately 700 µm² ($D_{void}$ = 30 µm). This value decreases when the shock reaches the void boundary and the void starts to collapse. Once the void has totally collapsed at t ≈ 11 ns, the $A_{void}$ value reaches zero. Due to the transformation of kinetic energy to thermal energy, a rapid increase in temperature is seen in Figure 4(b) and a hotspot of area $A_{hs}$ is formed, as shown in Figure 4(c). The reaction is initiated once a sufficiently strong hotspot is formed, leading to accumulation of reaction products. The time evolution of the product mass fraction $F$ is shown in Figure 4(d).

As noted in Eqn. (17) above, the QoI from the meso-scale is the time derivative $\dot{F}$ of the reacted mass fraction. By definition, $F$ is the ratio of the mass of reaction products ($M_{reacted}$) to the mass of HMX that would nominally fill the void ($M_{void}$):

$$F = \frac{M_{reacted}}{M_{void}} = \frac{\int_A \rho Y_4 \mathrm{d}A}{\rho_{HMX} A_{void}} \tag{21}$$

where, $\rho$ is the local density, $Y_4$ is the mass fraction of final gaseous species (Eqn. (10)), $\rho_{HMX}$ is the density of solid HMX and $A_{void}$ is the volume of the void. This way of defining the mass fraction of reaction products in a hotspot is similar to the one in Bastea et al. [45] who also tracked the evolution of the hotspot parameters after void collapse.

The product mass fraction evolution rate is:

$$\dot{F} = \frac{1}{M_{void}} \frac{dM_{reacted}}{dt} \tag{22}$$

Figure 4 shows the procedure to calculate $\dot{F}^{scv}_{ignition}$ and $\dot{F}^{scv}_{growth}$ for a single circular void. First, the variation of void area with time is plotted (Figure 3(a)) to locate two points in time. The first point (instant 1 in Figure (4)) is when $A_{void} = 90\%$ of its original (undeformed) area. The second point (instant 2 in the figure) is when the void area equals zero – i.e., when the void has totally collapsed. The various stages of collapse of a void, marking these points 1 and 2 can also be seen in Figure 4. Points 1 and 2 demarcate the ignition phase. The difference between the corresponding values of the product mass fraction $F$ and the time values for those two points are used to find the ignition rate $\dot{F}^{scv}_{ignition}$ as:

$$\dot{F}^{scv}_{ignition} = \frac{F_2 - F_1}{t_2 - t_1} \tag{23}$$

The hotspot growth rate is calculated using points 3 and 4. Point 3 is when the hotspot area equals 1.8 its value at point 2 (recall that point 2 is when the void has totally collapsed and the hotspot is ignited). Point 4 is when the hotspot area reaches twice its value at point 2. The shapes of the hotspots at points 3 and 4 are also shown in Figure 4. The differences in $F$ and time values for these points (3 and 4) are used to calculate the growth rate $\dot{F}^{scv}_{growth}$ as,

$$\dot{F}^{scv}_{growth} = \frac{F_4 - F_3}{t_4 - t_3} \tag{24}$$

This procedure is applied for all combinations of $(P_s, \tau_s, D_{void})$ to obtain the training data for $\dot{F}^{scv}_{ignition}$ and $\dot{F}^{scv}_{growth}$ for constructing the surrogates. The selection of the points in the $P_s, \tau_s, D_{void}$ space for obtaining the training data is described next.

2.5.2. Selection of the input points in the parameter space for obtaining the QoIs

As shown in a previous convergence study [11], the size of the training data set for reliable predictions will depend on the machine-learning algorithm. Here, the MBKG method [10] is used for which the required input sample size is approximately $8^d$, where d is the dimension of parameter space. This amounts to 512 simulations, which are used to construct the surrogates for $\dot{F}^{scv}_{ignition}$ and $\dot{F}^{scv}_{growth}$ in the $P_s, \tau_s, D_{void}$ space. For training the MBKG method, mesoscale computations are performed at 8 points along each parameter dimension for $P_s, \tau_s,$ and $D_{void}$. The training points along the $P_s$ dimension are chosen by supplying the $U_p$ values from 350 to 1500 m/s, at intervals of 150 m/s. The choice of specifying $U_p$ over $P_s$ is motivated by flyer-plate experiments, where $U_p$, as opposed to $P_s$, is a controllable design parameter. Similar to $P_s$, 8 training points are also selected along the $\tau_s^*$ dimension, with $\tau_s^*$ ranging from 0.1 to 5.0, at increments of 0.7. Finally, the values of $D_{void}$ are selected to be 1, 15, 30, 45, 60, 75, 90, and 100 μm for performing the mesoscale computations. The locations of the input data points for $D_{void}$ = 1, 15, 30, and 100 μm are shown in Figure 5.

It is noteworthy that not all the combinations of $P_s, \tau_s,$ and $D_{void}$ lead to sustained hot-spot growth at the mesoscale. If the input pressure is not sufficiently strong, or if $\tau_s$ is small, a hot-spot is eventually quenched by diffusion. These sub-critical cases are indicated by blue dots in Figure 5; at these locations in the parameter space, $\dot{F}^{scv}_{ignition}$ and $\dot{F}^{scv}_{growth}$ are negligible, because the chemical reaction does not lead to sustained reaction product species ($Y_4$) formation. The super-critical cases, i.e., the values of $P_s, \tau_s,$ and $D_{void}$, which lead to sustained hot-spot growth, are marked in red in Figure 5. The MBKG model is trained using both the sub- and super-critical inputs points in Figure 5.

## 3. RESULTS AND DISCUSSION

The methods described in the previous section are used to construct surrogate models for the QoIs ($\dot{F}^{scv}_{ignition}$ and $\dot{F}^{scv}_{growth}$) by performing high-fidelity mesoscale computations spanning the ($P_s, \tau_s, D_{void}$) parameter space. In this section, the surrogates for $\dot{F}^{scv}_{ignition}$ and $\dot{F}^{scv}_{growth}$ are presented first. Following this, we present physical insights on void dynamics that explain the characteristics of the surrogate models.

### 3.1. The surrogate models for $\dot{F}^{scv}_{ignition}$ and $\dot{F}^{scv}_{growth}$ in the $P_s, \tau_s, D_{void}$ space

The surrogates for $\dot{F}^{scv}_{ignition}$ and $\dot{F}^{scv}_{growth}$ are shown in Figures 6 and 7 respectively, for $D_{void}$ = 1, 15, 30, and 100 µm. The points in parameter space at which meso-scale reactive void collapse calculations are performed to obtain the surrogate models are shown in Figure 5. Similar to Figure 5, the sub-/super-critical regions in the surrogates are indicated by the red/blue zones on the hypersurfaces in Figures 6 and 7. Key features of the surrogate models observed in the figures are listed below.

1. Figures 5 through 7 show that there are jumps in criticality in the surrogates, i.e., there are sub-critical cases (blue dots) located between the critical/super-critical ones (red dots), This is seen, for example in Figure 5(c) and 6(c), for values of $\tau_s^* \geq 3$. This discontinuity in criticality is absent for the 1 µm case. There is only one region of discontinuity in the 15 µm surrogates in Figure 6, in the region of low $\tau_s^*$ (~0.1). For larger void diameters, for example, $D_{void}$= 30 µm and 100 µm, two regions are observed where the criticality condition is non-monotonic. The first is in the same $\tau_s^*$ (=0.1) region as for 15 µm void; the second region is for high values of $\tau_s^*$ (>3).
2. It is observed that $\dot{F}^{scv}_{ignition}$ and $\dot{F}^{scv}_{growth}$ for all void diameters reach their maximum values at $\tau_s^* \approx 1$, as seen in Figures 6 and 7. Thereafter, they decrease slightly and begin to saturate at $\tau_s^* \approx 2$. The saturation behavior of $\dot{F}^{scv}_{ignition}$ and $\dot{F}^{scv}_{growth}$ is observed in Figures 6(c)-(d) and Figures 7(c)-(d), with respect to $P_s$ as well, but only for large void diameters ($D_{void}$ = 30 and 100 µm). Thus, $\dot{F}^{scv}_{ignition}$ and $\dot{F}^{scv}_{growth}$ do not increase indefinitely with $\tau_s^*$ and $P_s$, but reach saturation for higher pressures as well as higher shock pulse thicknesses.
3. There is a non-monotonic trend of $\dot{F}^{scv}_{growth}$ with respect to $D_{void}$. As seen in Figure 7, the $\dot{F}^{scv}_{growth}$ for 15 µm and 30 µm voids is smaller than that for the 1 µm and 100 µm voids by an order of magnitude. In other words, $\dot{F}^{scv}_{growth}$ decreases with $D_{void}$, when 1 µm < $D_{void}$ < 30 µm, and increases thereafter, indicating that smaller voids (about 1 µm diameter) and large voids (>50 µm diameter) are more sensitive than voids of intermediate size. This type of trend reversal in sensitivity with void size has also been noted by Massoni et al. [46].

These three observations, their underlying physics and implications are discussed further in the following.

### 3.2. Criticality conditions for collapse of an isolated void

As shown in Figures 5 through 7, the criticality conditions with respect to $P_s$ and $\tau_s^*$ are non-monotonic, i.e., there are windows of subcritical collapses sandwiched between super-critical ones. To investigate this non-monotonic criticality behavior, three points in the parameter space are selected. These points are located in the regions of the jumps in criticality. The cases are for constant $D_{\text{void}}$ (=100 μm) and $\tau_s^*$ (= 5.0, i.e., approximating a sustained shock condition), and $P_s$ values of 2.8, 3.7, and 4.8 GPa. According to Figure 5, the case with $P_s$ = 3.7 GPa is sub-critical whereas the 2.8 GPa and 4.8 GPa cases are supercritical. The key question is: why does the void subject to the $P_s$ = 3.7 GPa fail to go critical, while the lower shock pressure of 2.8 GPa leads to criticality?

There are two primary aspects that contribute to criticality: (a) the mode of void collapse, which accounts for the hotspot shape, and (b) the intensity of void collapse, which accounts for hotspot temperature and size. Void collapse can occur in the plasticity-dominated mode or the hydrodynamic mode. The mode of collapse is determined by the ratio of the local pressure ($P_l$) around the void surface to the material yield strength ($Y$) [47]. When $P_l/Y \cong 1$, the void collapses in the plastic mode; whereas when the shock pressure overwhelms material strength, the void collapse mechanism switches to the hydrodynamic jetting mode. As will be shown below, the non-monotonic criticality with respect to $P_s$ is because the collapse mechanism shifts from plastic to a hydrodynamic mode as $P_s$ increases from 2.8GPa to 4.8GPa.

The mode in which the collapse occurs can be visualized from the velocity field accompanying the collapse event. Figure 8 shows the velocity contours at three different instants of time for the cases with $P_s$ = 2.8, 3.7, and 4.8 GPa in Figure 8(a), 8(b), and 8(c) respectively. As shown in Figure 8(a), for the lowest pressure, ($P_s$ = 2.8 GPa), the void collapses without the formation of a jet. This collapse is in the plasticity-dominated regime. The velocity magnitude in this case is smaller than the two higher pressure cases shown in Figures 8(b) and (c). The resulting hot-spot in this case has a circular spot shape, which is shown in Figure 9(a, II). This hotspot then grows over time and criticality is reached, as shown in Figure 9(a,III). Therefore, the $P_s$ = 2.8 GPa case leads to criticality, i.e., sustained growth of the hot-spot.

For the intermediate shock strength, i.e., for $P_s$ = 3.7GPa, the velocity field, even in the early stage (Figure 9(b),I) shows the formation of a weak jet. The jet then collapses the void, leading in Figure 9(b,III) to an arc-shaped hotspot. Figure 9(b) shows that this regime of transition from the plastic mode to jetting mode leads to a rather weak jet impact. The resulting arc-shaped hotspot is not intense enough to reach criticality and the hotspot fails to ignite, as shown in Figure 9(b,III).

For higher pressures, i.e., $P_s$ = 4.8GPa, the jet is stronger than in the intermediate shock strength case. The hotspot is formed due to jet impact, followed by the collapse of side lobes, as can be seen in Figure 9(c). The side-lobe collapse leads to very high temperatures, as shown in previous work [30] and also in the experiments of Bourne et al. [48], leading to an intense hotspot that is super-critical. Therefore, the case with highest pressure, $P_s$ = 4.8 GPa, goes critical, as shown in Figure 9(c III).

To summarize, the non-monotonic trend in the criticality of the hotspots is due to the switch in collapse mode from the plasticity-dominated collapse to the hydrodynamic jetting mode. At the intermediate pressure, which yields a sub-critical collapse, the jet is just being formed; the weak jet is unable to create a hotspot of sufficient intensity to sustain hot-spot growth. Therefore, a sub-critical case is sandwiched between the super-critical cases in Figures 5 and 6, leading to the non-monotonic criticality behavior in Figures 5 through 7.

The sweep of the 3-parameter space ($P_s$, $\tau_s$, $D_{\text{void}}$) is performed to demarcate points in the space that correspond to super- and sub-critical void collapse. The boundary separating these regions is the criticality envelope. Projection of this envelope on the $P_s - \tau_s$ and $P_s - D_{\text{void}}$ planes allows for the construction of scaling relationships for criticality for single isolated voids. Two scaling relationships are presented in this paper. The first envelope correlates the critical shock pressure and the critical pulse duration in the form:

$$\left(P_{s,\text{cr}}\right)^a \left(\tau_{s,\text{cr}}\right)^b = k \tag{25}$$

where $P_{s,cr}$ is the critical shock pressure, $\tau_{s,cr}$ is the corresponding critical shock pulse duration, and a, b, and k are constants to be determined. The constants a, b, and k are found using least-squares curve-fits to the critical pressures and pulse durations for various values of the void size $D_{void}$. The critical pressures for the various shock pulse durations are shown in the $P_s - \tau_s$ space in Figure 10(a)-(c) for $D_{void} = 1$, 15 and 30 μm. The diamond-shaped dots indicate criticality in the jetting collapse regime, while the circles indicate criticality in the plastic collapse regime. The least-squares curve fits are also shown in Figure 10. The constants a, b, and k are found to be: 4.09, 1, and 23.53 respectively, and the criticality criterion is therefore:

$$P_{s,cr}^{4.09} \tau_{s,cr} = 23.53 \qquad (26)$$

It is noteworthy that the relationship in Eqn. (26) is different from the Walker-Wasley relationship $P_{s,cr}^2 \tau_{s,cr} = $ constant or the related James' critical energy criterion [49] which applies at the macro-scale. In the multi-scale simulations of Sen et al. [8] it was shown that the Walker-Wasley type $P_{s,cr}^2 \tau_{s,cr} = $ constant relationship is indeed recovered by the MES-IG model for macro-scale SDT calculations using surrogate models for $\dot{F}_{ignition}^{scv}$ and $\dot{F}_{growth}^{scv}$ as obtained in this paper. At the macroscale, sub- or super-criticality is due to a competition between rarefaction, shocks and chemical reaction heat release in the material. On the other hand, at the mesoscale (i.e., in Eqn. (26)), the sub- or super-criticality results from a competition between thermal diffusion and the chemical reaction time-scales. For the collapse of a single void, the current $P_s - \tau_s$ relationship in Eqn. (26) is the meso-scale counterpart to the Walker-Wasley relationship $P_{s,cr}^2 \tau_{s,cr} = $ constant. Unlike the Walker-Wasley criterion [49], the relationship in Eqn. (26) does not carry the units of energy and is therefore a purely curve-fit scaling relationship. It nevertheless seems to apply for the range of the parameter space ($P_s$, $\tau_s$, $D_{void}$) examined in this paper. Eqn. (26) indicates that the meso-scale criticality criterion for a shock of given strength $P_s$ and pulse duration $\tau_s$ is strongly dependent on the pressure and only weakly on the pulse duration. In fact, as argued below, at the meso-scale, i.e., from the point of view of a single void, the pulse durations of shocks due to flyer impact are such that the void sees an approximation to a sustained pulse loading.

It is observed that the relationship in Eqn. (26) applies across the range of voids diameters tested. To see this, Figure 11 plots the quantity $P_{s,cr}^a$ against $\tau_{s,scr}^b$ for three different void diameters, viz. $D_{void} = $ 1μm, 15 μm, 30 μm. As expected, the figure shows that all values for all voids collapse onto a straight-line trend. Furthermore, the figure shows that the criticality regions overlap for different voids diameters. It is noted from the figure that smaller voids require higher shock pressure values to reach criticality but need smaller shock pulse durations $\tau_s$. Voids of different diameters occupy different regions of the criticality envelope, but all voids tested so far conform to the relationship in Eqn. (26).

From Eqn. (26) and Figure 9, it is also evident that criticality at the meso-scale depends strongly on the shock pressure $P_s$. For large $\tau_s^*$, the critical pressures saturate, i.e., the critical pressure becomes nearly independent of the pulse thickness. In most practical applications, a macro-scale sample of HMX is subjected to a shock pulse that, at the meso-scale, corresponds to $\tau_s^* > 1$. In that limit, an individual isolated void collapses as if it were impinged upon by a sustained shock pulse. Fixing the $\tau_s^*$ value at 5, i.e., for a pulse thick enough to be considered a sustained pulse, the data shown in Figure 5 can be used to develop a criticality relationship between $P_s$ and $D_{void}$. The criticality relationship is a fit of the form:

$$(P_{s,cr})^g (D_{void,cr})^h = q \qquad (27)$$

where $P_{s,cr}$ is the critical shock pressure, $D_{void,cr}$ is the critical void diameter, and g, h, and q are constants to be determined from a least-squares curve fit. Figure 12 shows the curve fit to the critical data points. The constants g, h, q in this formula are found to be 4.38, 1.0, and 3933 respectively, so that:

$$P_{s,cr}^{4.38} D_{void,cr} = 3933 \qquad (28)$$

Eqn. (28) shows that as the void diameter increases, the required pressure for criticality decreases. Therefore, smaller voids require higher pressure (shock strengths) to go critical. However, the growth rates of smaller voids can be larger, as shown below. An expression similar to the above is obtained from scaling analysis for critical void collapse in Rai et al. [47]. In Rai et al. [47], it is shown that the above criticality relationship can be derived from semi-analytically, using physical arguments similar to those underlying the often-used critical hotspot criterion of Tarver et al. [29]. However, the Tarver hotspot criticality criterion is expressed in terms of the hotspot characteristics, i.e., $T_{hs}$ and $D_{hs}$, both of which are not experimental control parameters. $T_{hs}$ and $D_{hs}$ are also not uniquely related to specific experimental conditions. In other words, one can arrive at the same hotspot temperature $T_{hs}$ or size $D_{hs}$ for different loading conditions $P_s$ and void sizes $D_{void}$. It is noted that, in contrast, Eqn. (28) is cast in terms of $P_s$ and $D_{void}$ and therefore provides a criticality criterion that can be useful for design of a HMX-based energetic material.

### 3.3. Variation of $\dot{F}^{scv}_{ignition}$ and $\dot{F}^{scv}_{growth}$ with $P_s$, $\tau_s$ and $D_{void}$

The surrogate models for $\dot{F}^{scv}_{ignition}$ and $\dot{F}^{scv}_{growth}$ are obtained using high-fidelity simulations spanning the parameter space ($P_s$, $\tau_s$, $D_{void}$). As pointed out in Section 3.1, the surrogates for $\dot{F}^{scv}_{ignition}$ and $\dot{F}^{scv}_{growth}$ saturate for certain values of $D_{void}$. The saturation is observed along both the $P_s$, $\tau_s$ directions. However, while saturation in the $\tau_s$ direction is observed for all $D_{void}$, saturation in the $P_s$ direction is observed only for large void diameters. These trends are investigated by examining the $\dot{F}^{scv}_{ignition}$ and $\dot{F}^{scv}_{growth}$ for different values of $D_{void}$ by taking sections of the hypersurfaces in the $P_s$ and $\tau_s^*$ directions. These are shown in Figures 13 and 14 for $\dot{F}^{scv}_{ignition}$ and $\dot{F}^{scv}_{growth}$ for $\tau_s^* = 5.0$.

It can be seen from Figure 13(a) that $\dot{F}^{scv}_{ignition}$ first increases for 1μm < $D_{void}$ < 30μm, then decreases until $D_{void}$ = 90μm and finally saturates as $D_{void}$ approaches 100 μm. Figure 14(a) shows that this non-monotonic behavior applies to $\dot{F}^{scv}_{growth}$ as well. This non-monotonicity is also clear from the plots in Figures 13(b) and 14(b), where $\dot{F}^{scv}_{ignition}$ and $\dot{F}^{scv}_{growth}$ are plotted as functions of $\tau_s^*$ alone, at a fixed $P_s = 22.1$ GPa. Furthermore, it is clear from Figures 12(b) and 13(b) that $\dot{F}^{scv}_{ignition}$ and $\dot{F}^{scv}_{growth}$ saturate quickly with the shock pulse duration. In general, saturation is achieved for $\tau_s^* > 1$. This is because shock pulse durations that are commensurate with the time of collapse of voids are most efficient at localizing shock energy at hotspots. Smaller pulse thicknesses ($\tau_s^* < 1$) supply insufficient energy to induce the collapse of voids, while larger pulse thicknesses ($\tau_s^* \gg 1$) do not contribute efficiently to enhancing the hotspot intensity. For long pulse durations, since the collapse ensues early in the shock passage time, the remaining compression due to the incident shock only heats up the post-collapse material to a modest degree -- most of the hotspot intensity is derived from the high temperatures resulting from the actual collapse event. Because the expected shock pulse thicknesses in flyer impact experiments are larger than the shock passage time over a single void, i.e., typically $\tau_s^* \gg 1$, most voids experience thick pulses and therefore the $\dot{F}^{scv}_{ignition}$ and $\dot{F}^{scv}_{growth}$ can be assumed to be nearly independent of the shock pulse durations in future efforts to construct surrogate models from meso-scale computations. This reduces the complexity of the surrogate model. Instead of surrogates in a three-dimensional parameter space ($\dot{F}^{scv}_{ignition}(P_s, \tau_s, D_{void})$ and $\dot{F}^{scv}_{growth}(P_s, \tau_s, D_{void})$), only two-dimensional parameter spaces can be used to compute $\dot{F}^{scv}_{ignition}(P_s, D_{void})$ and $\dot{F}^{scv}_{growth}(P_s, D_{void})$ for the MES-IG model, thereby reducing the computational cost for constructing the surrogates. The use of $\dot{F}^{scv}_{ignition}(P_s, D_{void})$ and $\dot{F}^{scv}_{growth}(P_s, D_{void})$ in macro-scale simulations is demonstrated in Sen et al. [8].

It is noted that the non-monotonic trends in the $\dot{F}^{scv}_{ignition}(P_s, D_{void})$ and $\dot{F}^{scv}_{growth}(P_s, D_{void})$ arise due to the normalization of $\frac{dM_{reacted}}{dt}$ with $M_{void}$ in Eqn. (21). In contrast to the normalized reaction rates, the rate of generation of reaction products, i.e., the quantity $\frac{dM_{reacted}}{dt}$ itself varies monotonically with both $P_s$ and $\tau_s$, as shown in Figures 13(c-d) and 14(c-d). This is because larger voids produce greater amount of reaction products over the entire range of pressure and shock durations, as would be expected. However, when normalized by $M_{void}$ to obtain the $\dot{F}^{scv}_{ignition}(P_s, D_{void})$ and $\dot{F}^{scv}_{growth}(P_s, D_{void})$, the non-monotonic behavior with respect to $D_{void}$ is obtained. The smaller voids of around 1 μm diameter therefore appear to be more efficient at localizing energy than mid-size voids (in the 30-45 μm diameter range). Similarly, larger voids, in the range of 60-100 μm diameter also appear to be efficient in localizing energy. However, such large voids are likely to be few and far between in typical meso-structures and may not contribute significantly to the porosity of the material [46]. Voids sizes centered around the 1 μm diameter range, which are likely more abundant, therefore are more likely to increase the sensitivity of a HMX-based HE material. It is noted that other researchers have also indicated that voids in the range of 0.1 μm-10 μm are most sensitive [28, 50].

It is noticed from Figure 6(a) and (d) as well as Figure 7(a) and (d) that $\dot{F}^{scv}_{ignition}(P_s, D_{void})$ and $\dot{F}^{scv}_{growth}(P_s, D_{void})$ saturate with respect to $P_s$ for the larger diameter (e.g. 10 μm) void but not for the smaller diameter (1 μm) void. To understand why saturation occurs for the larger voids as the pressure increases, Figure 15 plots the evolution of hotspot shapes for a void of 10 μm diameter as the pressure is increased from 1.27 GPa (plastic collapse regime, with a circular hotspot upon collapse) to 28.9 GPa (hydrodynamic regime, with a mushroom-shaped hotspot). For the larger void ($D_{void} = 10$ μm), as the pressure is increased, the hotspot initially changes shape, size and temperature. This is seen between $P_s = 1.27$ GPa in Figure 15(a) to $P_s = 18.5$ GPa in Figure 15(g). Thereafter, all the hotpot metrics, i.e., shape, size and temperature appear to saturate. This results in the saturation of $\dot{F}^{scv}_{ignition}(P_s, D_{void})$ and $\dot{F}^{scv}_{growth}(P_s, D_{void})$ for $D_{void} = 10$ μm.

In contrast with $D_{void} = 10$μm, for the smaller voids (e.g. $D_{void} = 1$μm), $\dot{F}^{scv}_{ignition}(P_s, D_{void})$ and $\dot{F}^{scv}_{growth}(P_s, D_{void})$ do not saturate with respect to $P_s$. Figure 16 contrasts the situation for $D_{void} = 1$μm, with that of Fig 15 ($D_{void} = 10$ μm). Unlike the 10 μm void, for the 1μm void, the hotspot metrics change across the range of pressures $P_s$ used in the collapse simulations. The hotspot shape changes even in the high range of pressures, i.e., between $P_s = 22.1$ GPa and $P_s = 54.3$ GPa. The $\dot{F}^{scv}_{ignition}(P_s, D_{void})$ and $\dot{F}^{scv}_{growth}(P_s, D_{void})$, which are measures of the effectiveness of energy localization in a hotspot (see Eqn. (21)), follow the change in the primary metrics of the hotspot and do not saturate with increasing shock strength ($P_s$ and $\tau_s$) for $D_{void} = 1$μm.

Figures 17 and 18 show the functions $\dot{F}^{scv}_{ignition}(P_s, D_{void})$ and $\dot{F}^{scv}_{growth}(P_s, D_{void})$ in the $P_s - D_{void}$ space, for a constant value of $\tau_s^* = 5$. From Figure 17 it is seen that $\dot{F}^{scv}_{ignition}$ depends strongly on pressure but only weakly on the void size. This is aligned with the fact that the probability of ignition of a hotspot depends on the intensity of collapse, i.e., the temperature achieved upon collapse. While there appears to be some variation of $\dot{F}^{scv}_{ignition}$ with respect to $D_{void}$ at high pressures, the variation with respect to $D_{void}$ in the bulk of the pressure range of 1-20 GPa is small. In contrast, the $\dot{F}^{scv}_{growth}$ in Figure 17 shows a strong dependence on the void size. This is because, once the void collapses, further growth of the hotspot depends on the competition between chemical reaction, thermal diffusion and transport of heat by advection. Of these, chemical reaction front growth and thermal diffusion depend on the size of the hotspot. In Rai et al. [47] the scaling laws and non-dimensional parameters that control the interplay between chemistry, diffusion and advection are obtained to explain the observed shape of the surface $\dot{F}^{scv}_{growth}(P_s, D_{void})$. In particular the non-monotonic behavior of $\dot{F}^{scv}_{growth}$ is noteworthy. In particular, at high pressures, $\dot{F}^{scv}_{growth}$ first decreases sharply from its value at $D_{void} = 1$ μm to about $D_{void} = 30$ μm. Thereafter $\dot{F}^{scv}_{growth}$ increases again to saturate at around $D_{void} = 100$ μm.

In related work of Rai et al. [47], using scaling arguments this nonlinear variation of $\dot{F}^{scv}_{growth}$ is investigated further and explained. Briefly, in the range $D_{void} = 1$ μm to 30μm, the decrease in $\dot{F}^{scv}_{growth}$ is due to the change in the shape of the hotspot. At low values of $D_{void}$, the growth rate is increased by advection and stretching of hotspot surfaces because convection (which is independent of hotspot size) outcompetes chemical reaction front growth (which is proportional to hotspot size and is therefore small for small hotspots). In the range of $D_{void} = 45$ μm to 100μm, the rate of growth of the reaction front is rapid enough that the chemical reaction front catches up with and feeds the imposed shock leading to the propagation of a burning front that accompanies and reinforces the propagating shock wave. Figure 19 shows the difference in the hotspot growth between the minimum value in Figure 17, i.e., $D_{void} = 30$μm and the point of transition to the "micro-detonation" mode at $D_{void} = 45$μm. As shown in Figure 19(a), for the case of $D_{void} = 30$μm the hotspot grows from a mushroom shape in Figure 19(a,II) to a nearly circular shape in Figure 19(a,IV). This nearly circular shape leads to a deceleration of the hotspot growth since a small reaction front surface results in Figure 19(a,IV). However, in the case of the $D_{void} = 45$ μm situation, the reaction front is seen to catch up with the incident shock in Figure 19(b,II). The subsequent evolution leads to a strong blast wave with a post-shock region that is fully burned material. The coupling between the reaction and detonation regions continues in this case to later times, as seen in Figure 19(b,IV) leading to high $\dot{F}^{scv}_{growth}$. Therefore, the growth of the hotspots, as represented by the quantity $\dot{F}^{scv}_{growth}$ depends strongly on both the loading ($P_s$) and the void size $D_{void}$ as can be observed in Figure 18. These dependencies of $\dot{F}^{scv}_{growth}$ on $D_{void}$ are transmitted to the macro-scale in the MES-IG model, as demonstrated in multi-scale simulations of Sen et al. [8], leading to the variation with $D_{void}$ of run-to-detonations and critical energies that define go-no-go envelopes.

## 4. CONCLUSIONS

This work presents the procedure for constructing a machine-learned surrogate model for hotspot ignition and growth rates in pressed HMX materials. A Bayesian Kriging algorithm is used to assimilate input data obtained from high-resolution meso-scale simulations. The surrogates are built by generating a sparse set of training data using reactive meso-scale simulations of void collapse with different loading conditions and varying void sizes. The information obtained from these surrogate models can be utilized in meso-informed Ignition and Growth (MES-IG) multi-scale simulations of pressed HMX materials. The first step in building the MES-IG model is to obtain the training or input data for the surrogate models:

$$\dot{F}_{ignition} = \dot{F}^{scv}_{ignition}(P_s, \tau_s, D_{void}) * f^{v-v}_{ignition}(\phi) * f^{shape}_{ignition}(AR, \theta)$$

and

$$\dot{F}_{growth} = \dot{F}^{scv}_{growth}(P_s, \tau_s, D_{void}) * f^{v-v}_{growth}(\phi) * f^{shape}_{growth}(AR, \theta)$$

where $\dot{F}^{scv}_{ignition}(P_s, \tau_s, D_{void})$, $\dot{F}^{scv}_{growth}(P_s, \tau_s, D_{void})$ consider the effect of loading and void size on a single cylindrical void, and $f^{v-v}_{ignition}(\phi), f^{v-v}_{growth}(\phi)$ and $f^{shape}_{ignition}(AR, \theta), f^{shape}_{growth}(AR, \theta)$ consider the void-void interaction and void shape effects respectively. The work in this paper is focused towards obtaining the functions $\dot{F}^{scv}_{ignition}(P_s, \tau_s, D_{void})$ and $\dot{F}^{scv}_{growth}(P_s, \tau_s, D_{void})$ for single circular voids.

The sweep of the 3-parameter space $(P_s, \tau_s, D_{void})$ provides points in the space that represent super- and sub-critical void collapse. The boundary separating these regions is the criticality envelope. Projection of this envelope on the $P_s - \tau_s$ and $P_s - D_{void}$ planes allows for the construction of scaling relationships for the critical conditions for single cylindrical voids. Two scaling relationships are presented in this paper:

$$P^{4.09}_{s,cr} \tau_{s,cr} = 23.53$$

This relationship is the meso-scale counterpart to the Walker-Wasley relationship $P_{s,cr}^2 \tau_{s,cr} = $ constant which applies at the macro-scale. In most practical applications, a macro-scale sample of HMX is subjected to a shock pulse that, at the meso-scale corresponds to thick shock pulses. In that limit, an individual isolated void collapses as if subject to a sustained shock pulse. Assuming thick shock pulses at the meso-scale, the surrogate models can be used to develop a criticality relationship between $P_s$ and $D_{\text{void}}$:

$$P_{s,cr}^{4.38} D_{\text{void,cr}} = 3933$$

This criticality relationship is useful because it is expressed in terms of the control parameters, i.e., shock strength $P_s$ and meso-structural parameter $D_{\text{void}}$. The above criticality relationship is derived from considerations similar to the often used critical hotspot criterion of Tarver et al. [29]. However, the latter is expressed in terms of the hotspot characteristics, i.e., $T_{\text{hs}}$ and $D_{\text{hs}}$, both of which are not experimental control parameters, i.e., one can arrive at a given hotspot temperature $T_{\text{hs}}$ or size $D_{\text{hs}}$ for different loading conditions $P_s$ and void sizes $D_{\text{void}}$. Therefore, the above criticality relationship is in a form that is of greater utility for design of HMX based HEs.

The surrogate models for $\dot{F}_{\text{ignition}}^{\text{scv}}$ and $\dot{F}_{\text{growth}}^{\text{scv}}$ are obtained using 512 simulations spanning the parameter space ($P_s, \tau_s, D_{\text{void}}$). There are some distinct trends in the $\dot{F}_{\text{ignition}}^{\text{scv}}$ and $\dot{F}_{\text{growth}}^{\text{scv}}$ observed from the surrogate model hypersurfaces. It is seen that $\dot{F}_{\text{ignition}}^{\text{scv}}$ depends strongly on pressure but only weakly on the void size. This is aligned with the fact that the probability of ignition of a hotspot depends on the intensity of collapse, i.e., the temperature achieved upon collapse. While there appears to be some variation of the function with respect to void diameter at high pressures, the variation with respect to $D_{\text{void}}$ in the bulk of the pressure range of 1-20 GPa is small. In contrast, the $\dot{F}_{\text{growth}}^{\text{scv}}$ shows a strong dependence on the void size. This is because, once the void collapses, the further growth of the hotspot depends on the competition between chemical reaction, thermal diffusion and transport of heat by advection. Of these, chemical reaction front growth and thermal diffusion depend on the size of the hotspot. Therefore, the growth of the hotspots, as represented by the quantity $\dot{F}_{\text{growth}}^{\text{scv}}$ depends strongly on both the loading ($P_s$) and the void size $D_{\text{void}}$. These dependencies on $D_{\text{void}}$ are transmitted to the macro-scale in the MES-IG model [8], leading to the variation of run-to-detonations and critical energies that define go-no-go envelopes in macro-scale samples.

The surrogate models developed in this paper encapsulate the meso-scale dynamics that contribute to the localization of energy due to shock-induced void collapse. The model construction process relies on simplified physical models of the material (HMX) and idealizations of the void shape (2D, cylindrical). Based on our previous work [31], it is expected that for collapse of 3D voids, the $\dot{F}_{\text{ignition}}^{\text{scv}}$ and $\dot{F}_{\text{growth}}^{\text{scv}}$ surrogates will be higher than the ones shown in the paper. Furthermore, the present surrogates also apply to an isolated void in an otherwise homogeneous material. In Part II the effects of void-void interaction in a field of voids and deviations from circular void shape are studied. Limitations of the present model, in terms of material models for HMX, chemical kinetics models etc, introduce as yet undetermined epistemic uncertainties in the surrogate models. The quantification of the uncertainties, their propagation to the macroscale and the effect on the predicted run-to-detonation response of HEs is the subject of ongoing work.

**ACKNOWLEDGEMENTS**

The authors gratefully acknowledge the financial support from the Air Force Office of Scientific Research (Dynamic Materials Program, program manager: Martin Schmidt) under grant number FA9550-15-1-0332 and EGLIN AFB, AFRL-RWPC (program manager: Angela Diggs) under the contract number FA8651-16-1-0005. The authors are also thankful to K.K. Choi at the University of Iowa and Nicholas J. Gaul at RAMDO LLC, Iowa City, for providing the computational code for the Modified Bayesian Kriging Method.

**FIGURES**

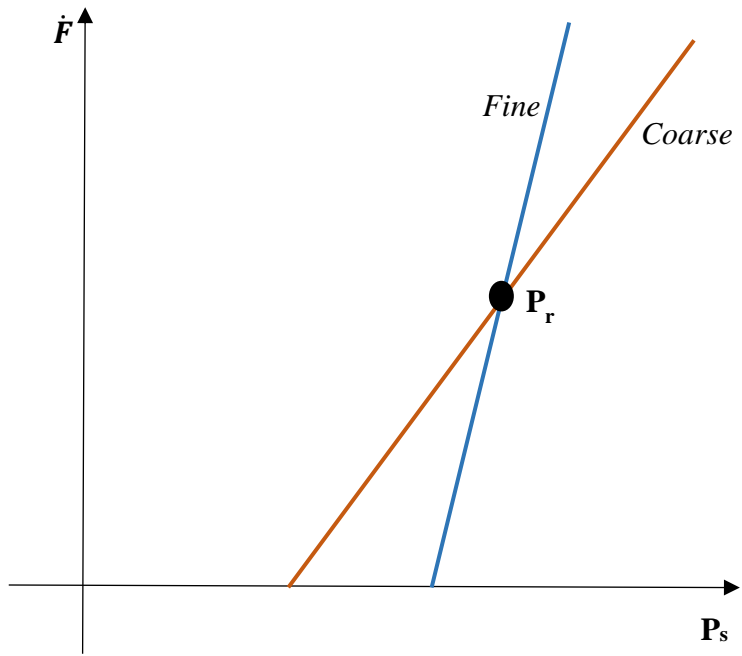

Figure 1. A schematic of the effect of particle size on the sensitivity of HE, reproduced from Price [12]. The coarse grained HE (represented by the red line) ignites more readily, i.e., at lower shock pressures but the build-up to detonation, i.e., the reaction rate is lower. The fine grain particles (indicated by the blue line) ignite at higher pressures, but once ignited the buildup to detonation is more rapid. The crossover pressure for coarse-to-fine sensitivity occurs at a pressure $P_r$.

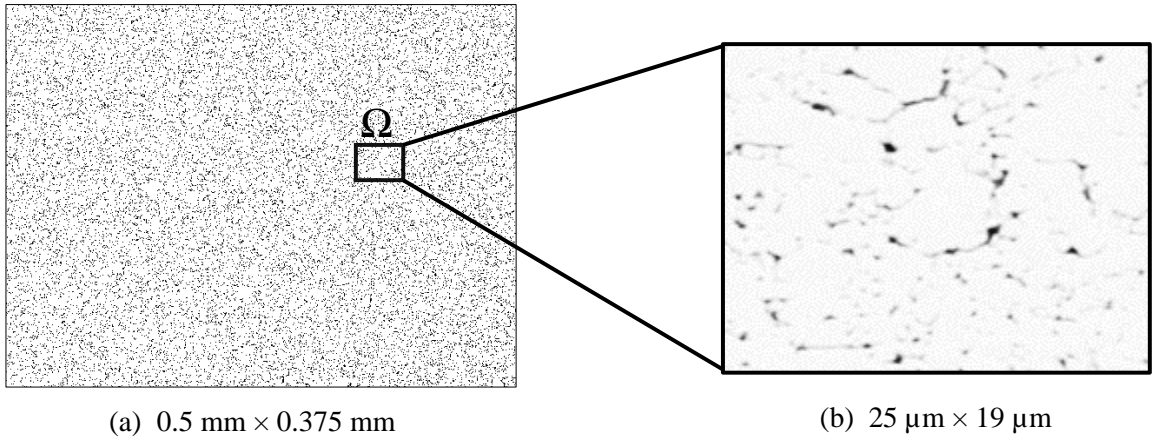

(a) 0.5 mm × 0.375 mm  (b) 25 µm × 19 µm

Figure 2: An SEM image [1] of a pressed HMX; the dimensions of the sample are 0.5 mm × 0.375 mm. The black dots in the image show the voids in the material. A sub-region of the sample, indicated by Ω, is zoomed in (b), showing the meso-structure of the sample. The dimension of the sub-sample is 25 µm × 19 µm.

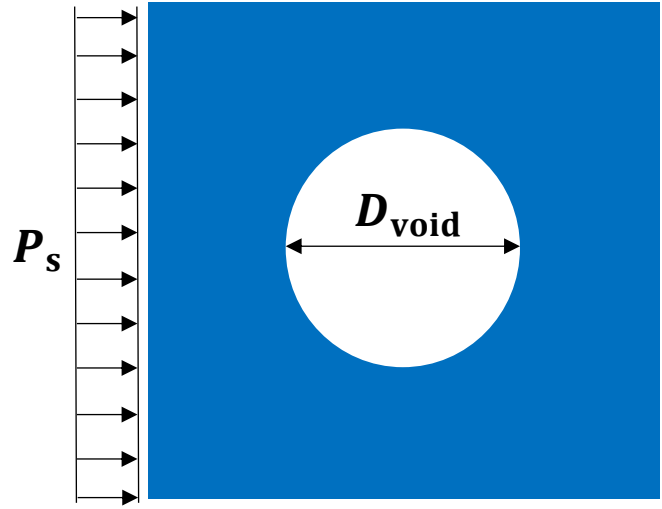

Figure 3: The numerical setup for performing high-fidelity mesoscale simulations of reactive void-collapse to construct the surrogate $\dot{F}^{\text{scv}}_{\text{ignition}}$ and $\dot{F}^{\text{scv}}_{\text{growth}}$. A single cylindrical void with diameter $D_{\text{void}}$ is subjected to a shock pressure $P_s$, with $\tau_s$ denoting the duration of the loading. The load is applied from the west side of the domain boundary, as shown in the figure.

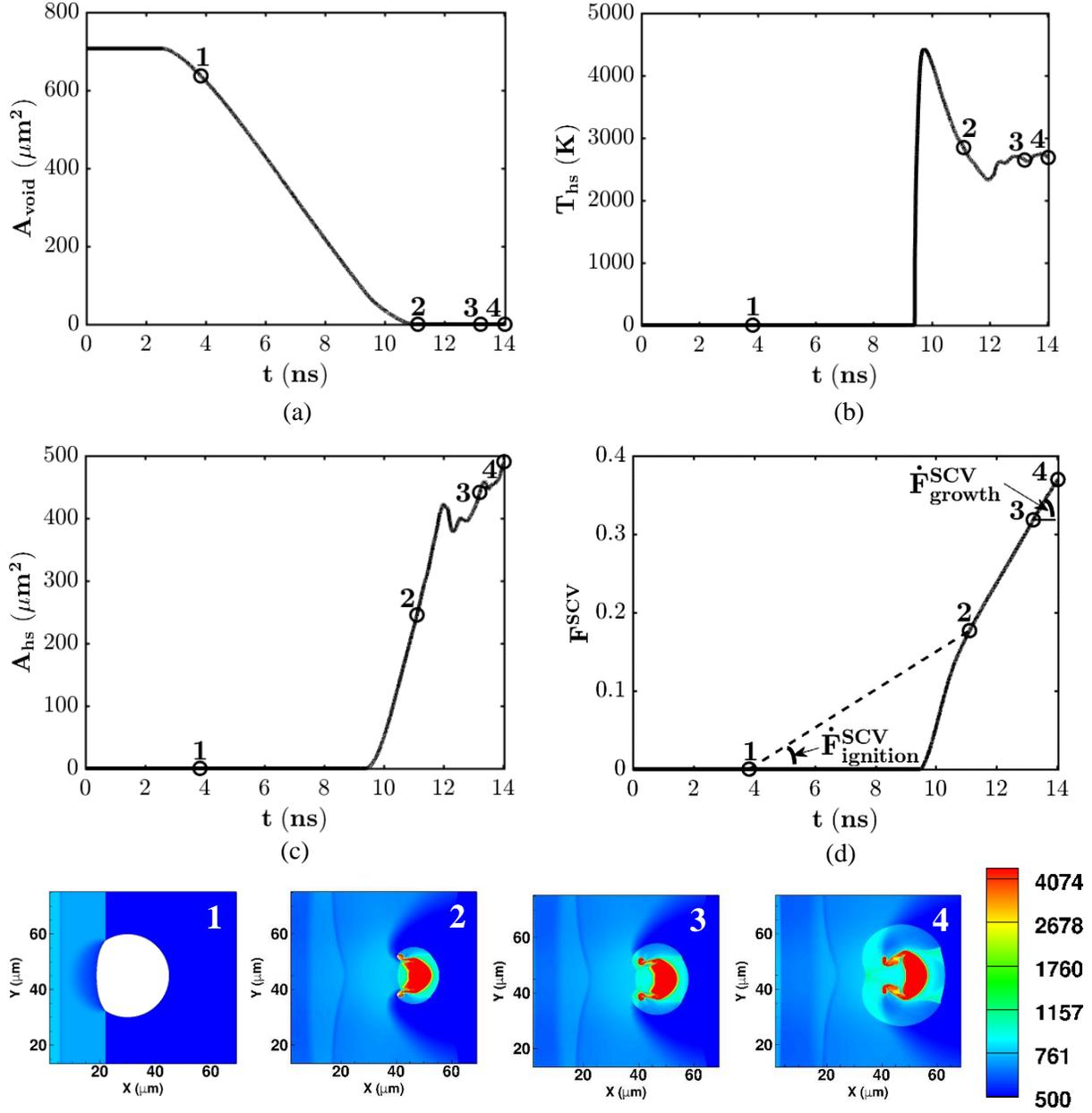

Figure 4: Illustration of the procedure for obtaining $\dot{F}^{scv}_{ignition}$ and $\dot{F}^{scv}_{growth}$ from the mesoscale numerical experiments. The illustration is shown for $D_{void} = 30$ µm, subjected to a shock pressure of $P_s = 22.1$ GPa and $\tau_s^* = 1.5$. The figures show the variation of (a) void area, (b) hotspot average temperature, (c) hotspot area, and (d) the products mass fraction $F$ (i.e., $\frac{M_{reacted}}{M_{pore}}$) with time. The numbers 1 through 4 in (a)-(d) indicate different instances of void and hot-spot formation/growth. In Figure (d), the time instances selected for computing the slopes to obtain $\dot{F}^{scv}_{ignition}$ and $\dot{F}^{scv}_{growth}$ are shown. The corresponding temperature contours (K) at instances 1-4 are also shown in the figure.

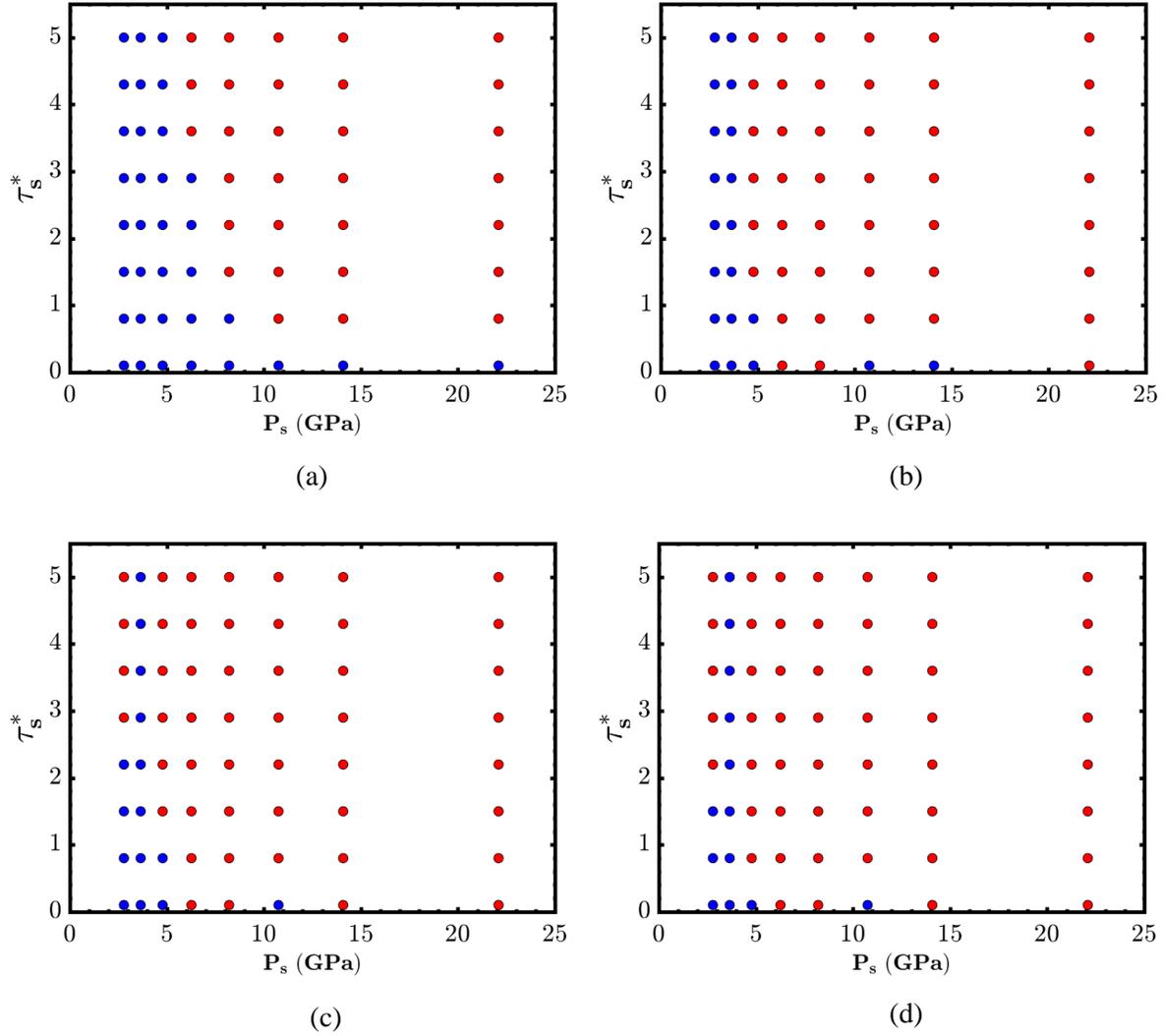

Figure 5: Location of the training points for constructing the surrogate models for $\dot{F}^{\text{scv}}_{\text{ignition}}$ and $\dot{F}^{\text{scv}}_{\text{growth}}$ in the $P_s$ - $\tau_s^*$ space for $D_{\text{void}}$ = (a) 1 μm, (b) 15 μm, (c) 30 μm and (d) 100 μm. The blue dots are the sub-critical cases (i.e., where the hot spot was quenched by diffusion), and the red dots are the critical/super-critical cases (i.e., where the hot-spot lead to sustained chemical reaction).

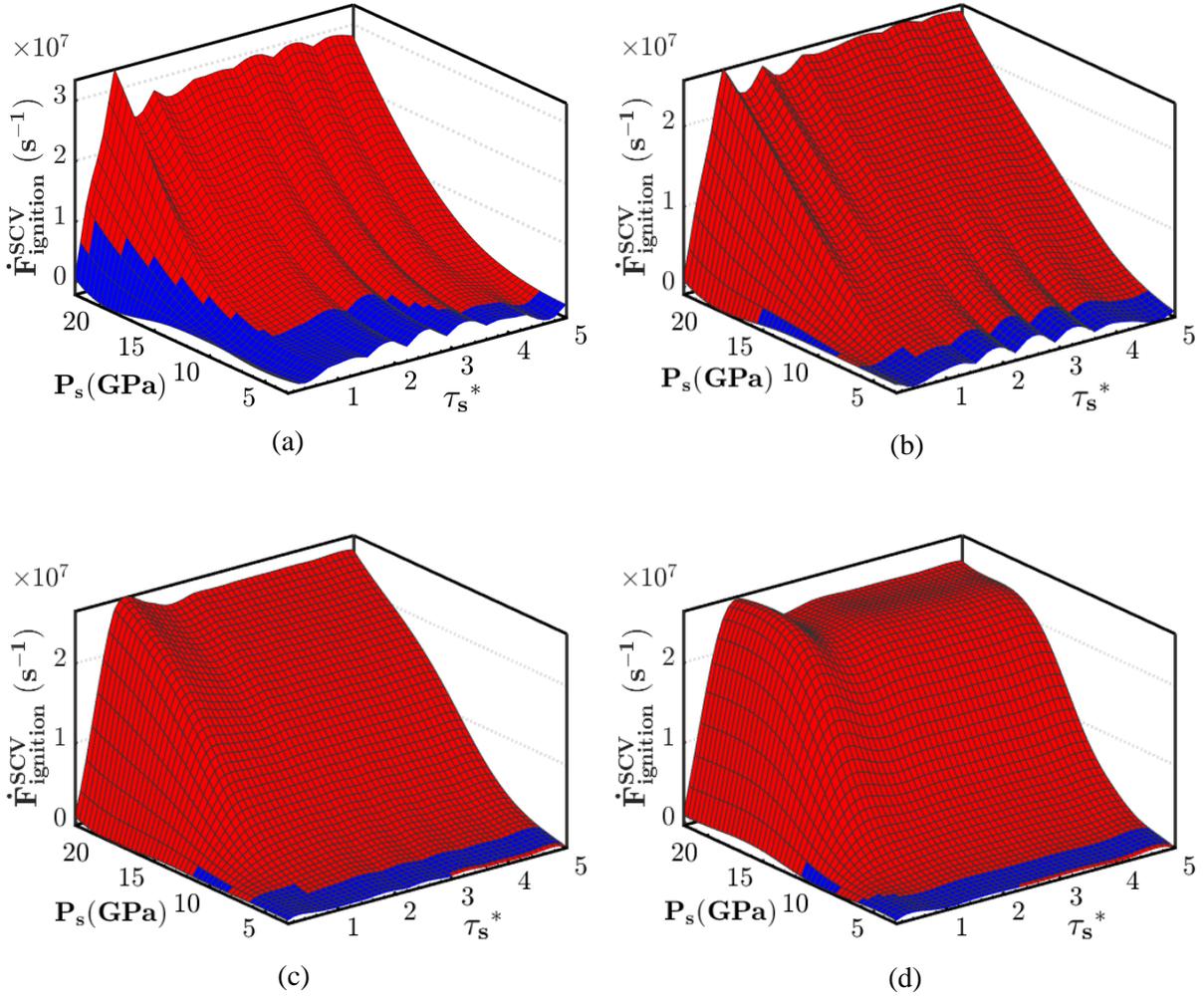

Figure 6: Surrogates of the ignition reaction rate $\dot{F}_{\text{ignition}}^{\text{scv}}$ showing its variation with respect to shock pressure $P_s$ and to the dimensionless shock duration $\tau_s^*$. The surrogates are shown for $D_{\text{void}}$ = (a) 1 μm, (b) 15 μm, (c) 30 μm and (d) 100 μm. The blue regions are the sub-critical zones (i.e., where the hot spot was quenched by diffusion), whereas the red ones are the critical/super-critical zones (i.e., where the hot-spot lead to sustained chemical reaction).

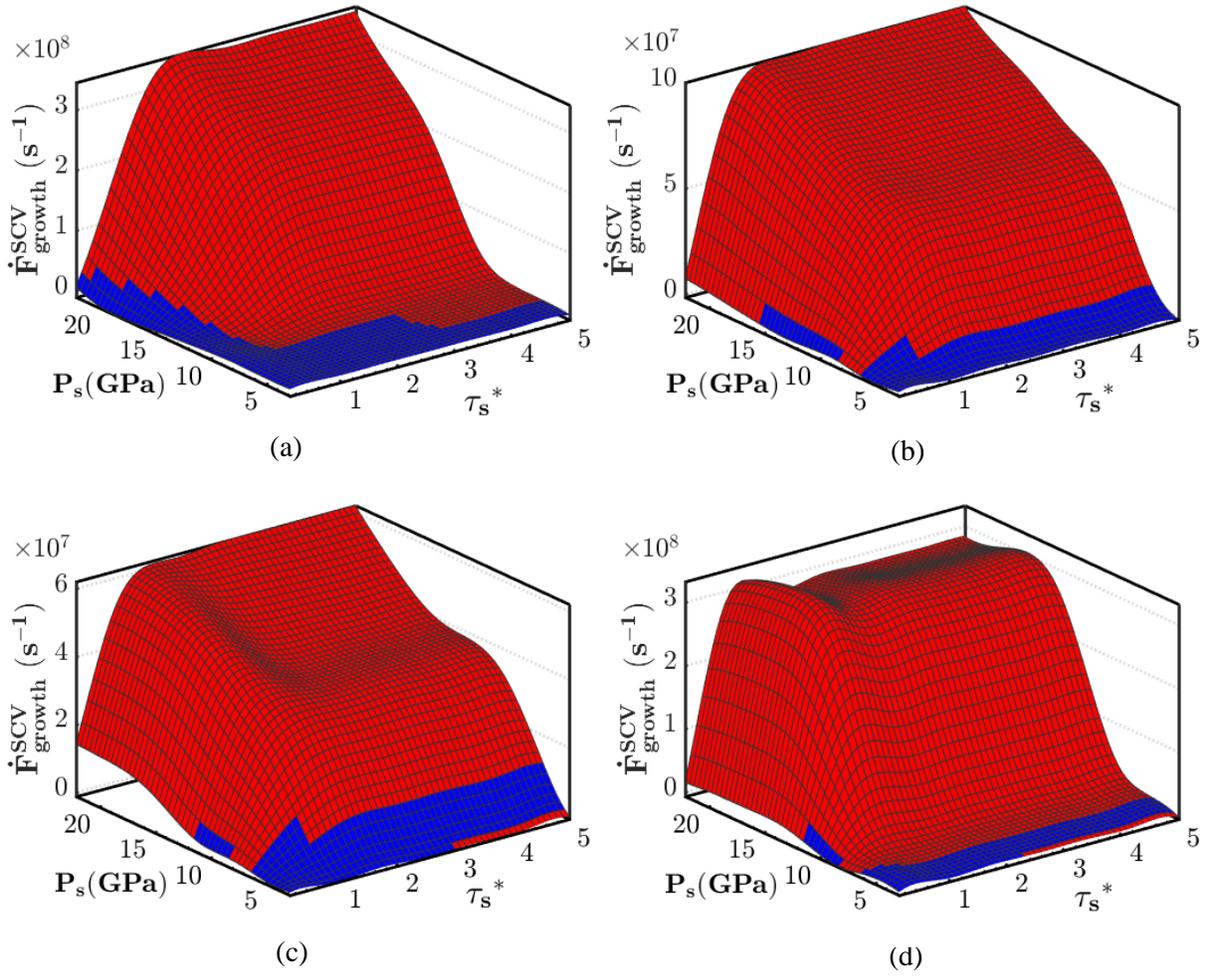

Figure 7: Surrogates of the ignition reaction rate $\dot{F}_{growth}^{scv}$ showing its variation with respect to shock pressure $P_s$ and to the dimensionless shock duration $\tau_s^*$. The surrogates are shown for $D_{void}$ = (a) 1 μm, (b) 15 μm, (c) 30 μm and (d) 100 μm. The blue regions are the sub-critical zones (i.e., where the hot spot was quenched by diffusion), whereas the red ones are the critical/super-critical zones (i.e., where the hot-spot lead to sustained chemical reaction).

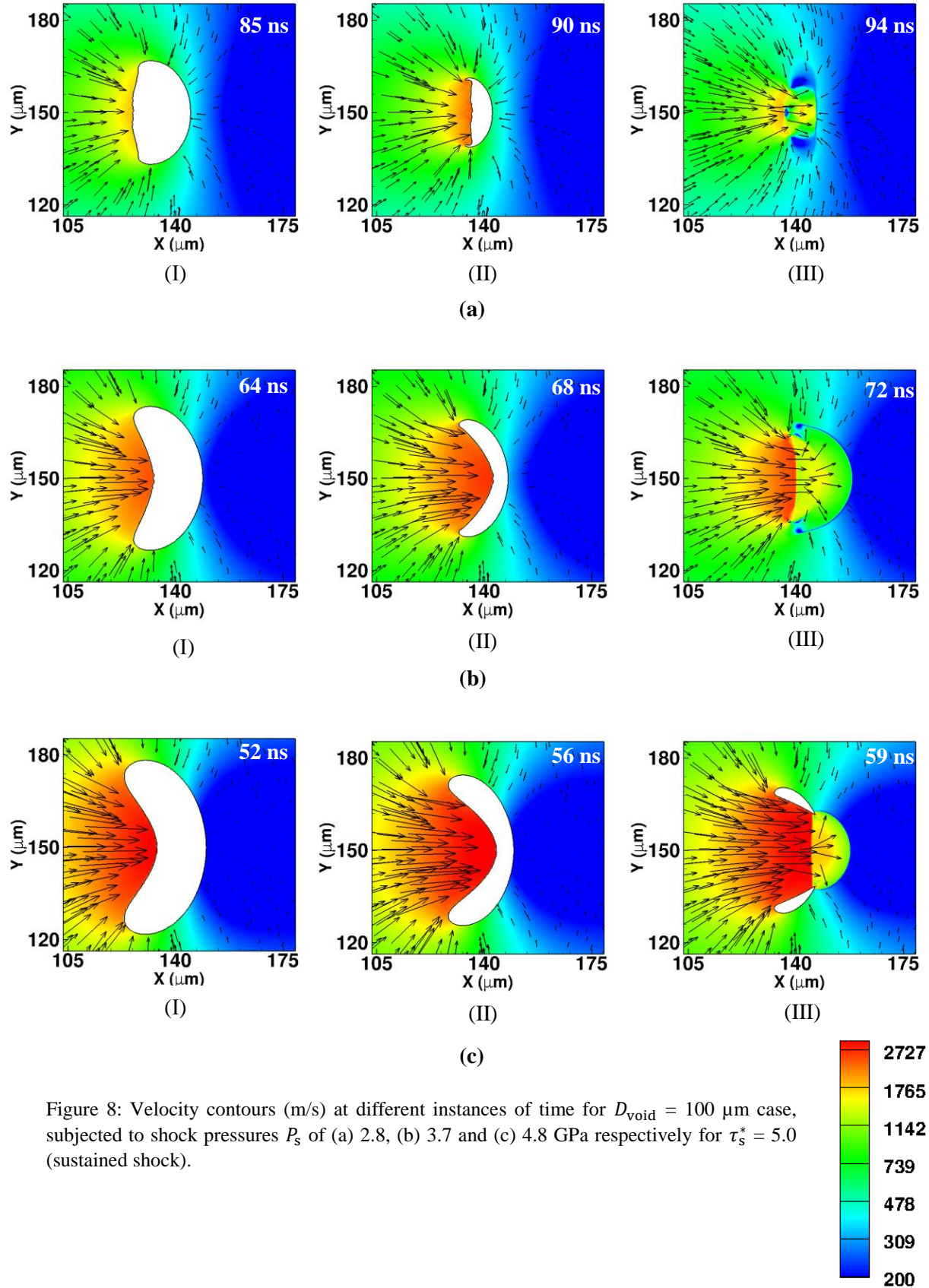

Figure 8: Velocity contours (m/s) at different instances of time for $D_{\text{void}} = 100$ µm case, subjected to shock pressures $P_s$ of (a) 2.8, (b) 3.7 and (c) 4.8 GPa respectively for $\tau_s^* = 5.0$ (sustained shock).

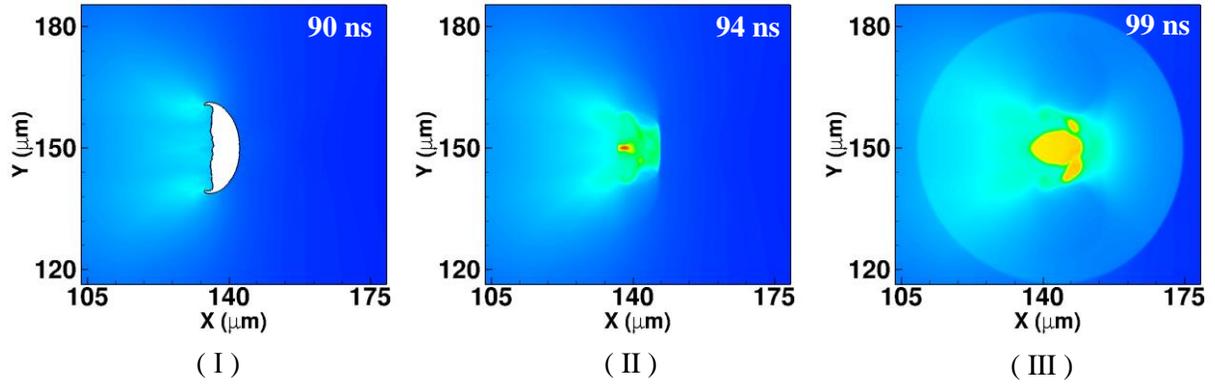

(a)

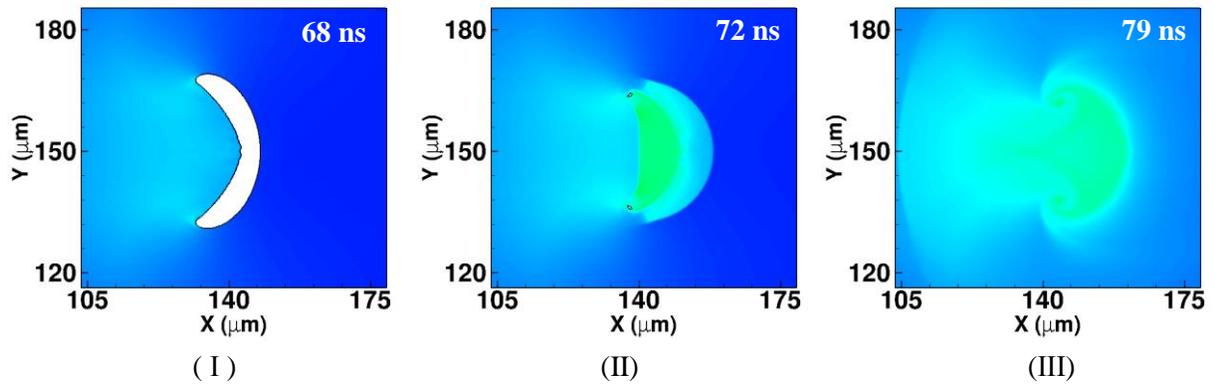

(b)

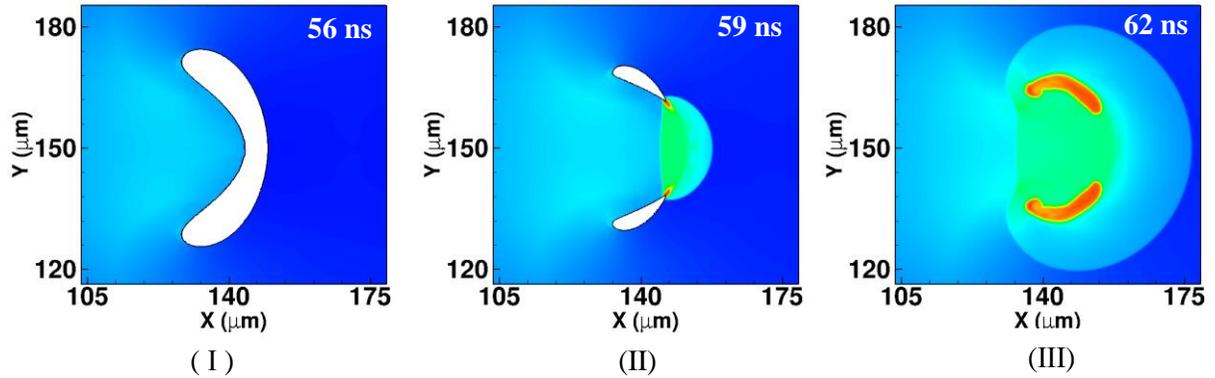

(c)

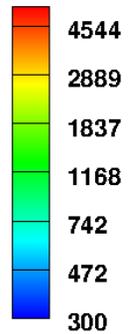

Figure 9: Temperature contours (K) at different instances of time for $D_{\text{void}} = 100$ μm case, subjected to shock pressures $P_s$ of (a) 2.8, (b) 3.7 and (c) 4.8 GPa respectively for $\tau_s^* = 5.0$ (sustained shock).

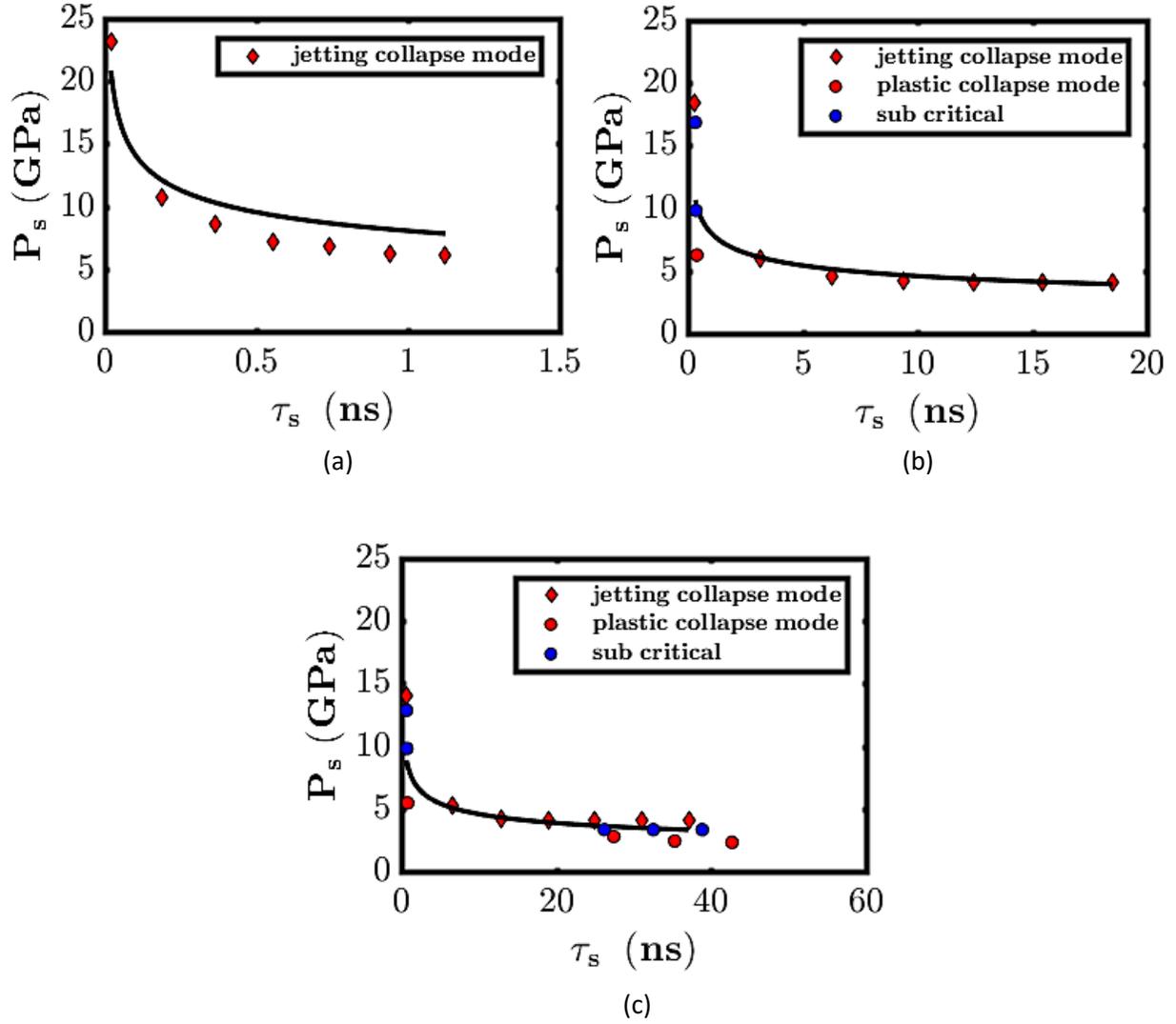

Figure 10: The critical shock pressure $P_{s,cr}$ vs. the critical shock duration $\tau_{s,cr}$ with curves fitted to the formula $(P_s)^a (\tau_s)^b = k$ for 1, 15, and 30 µm voids diameters in figures (a-c) respectively. The constants a, b, and k are found to be 4.09, 1, 23.53 respectively. The red dots in the figure are the critical/super-critical points where the hot-spot due to void collapse led to a sustained hot-spot growth. The blue dots are the sub-critical points, where the hot-spot is eventually quenched by diffusion.

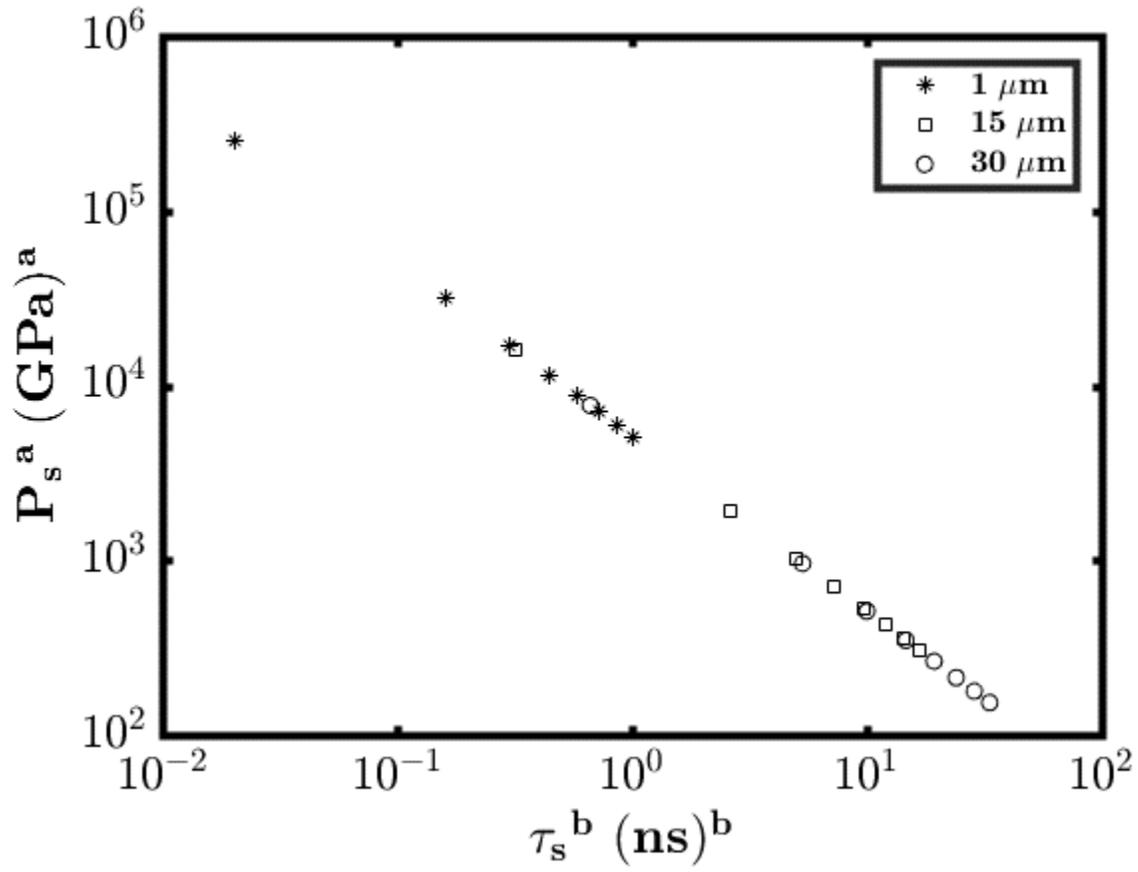

Figure 11: Plot for the critical points for 1, 15, and 30 µm voids diameters plotted based on the scaling law, $P_s^{4.09}\tau_s = 23.53$.

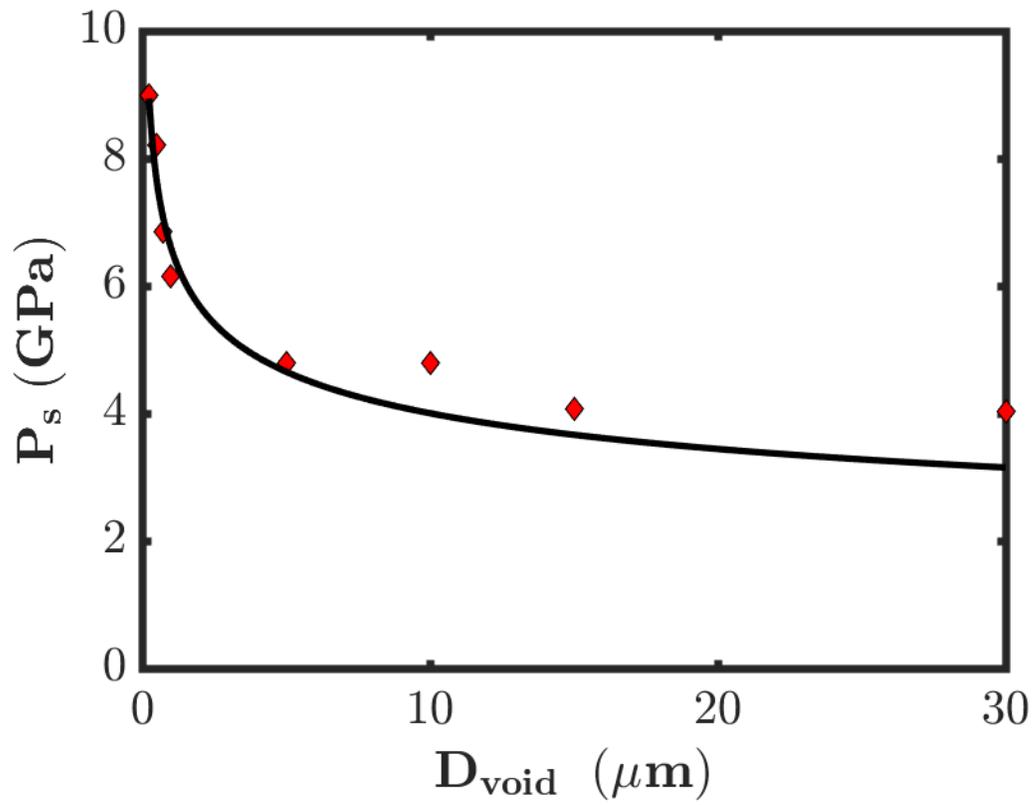

Figure 12: Critical shock pressure $P_s$ vs. void diameter for $\tau_s^* = 5.0$ with a curve fitted to the formula $(P_s)^g (D_{void})^h = q$. The constants g, h, q in this formula are 4.38, 1.0, and 3933 respectively.

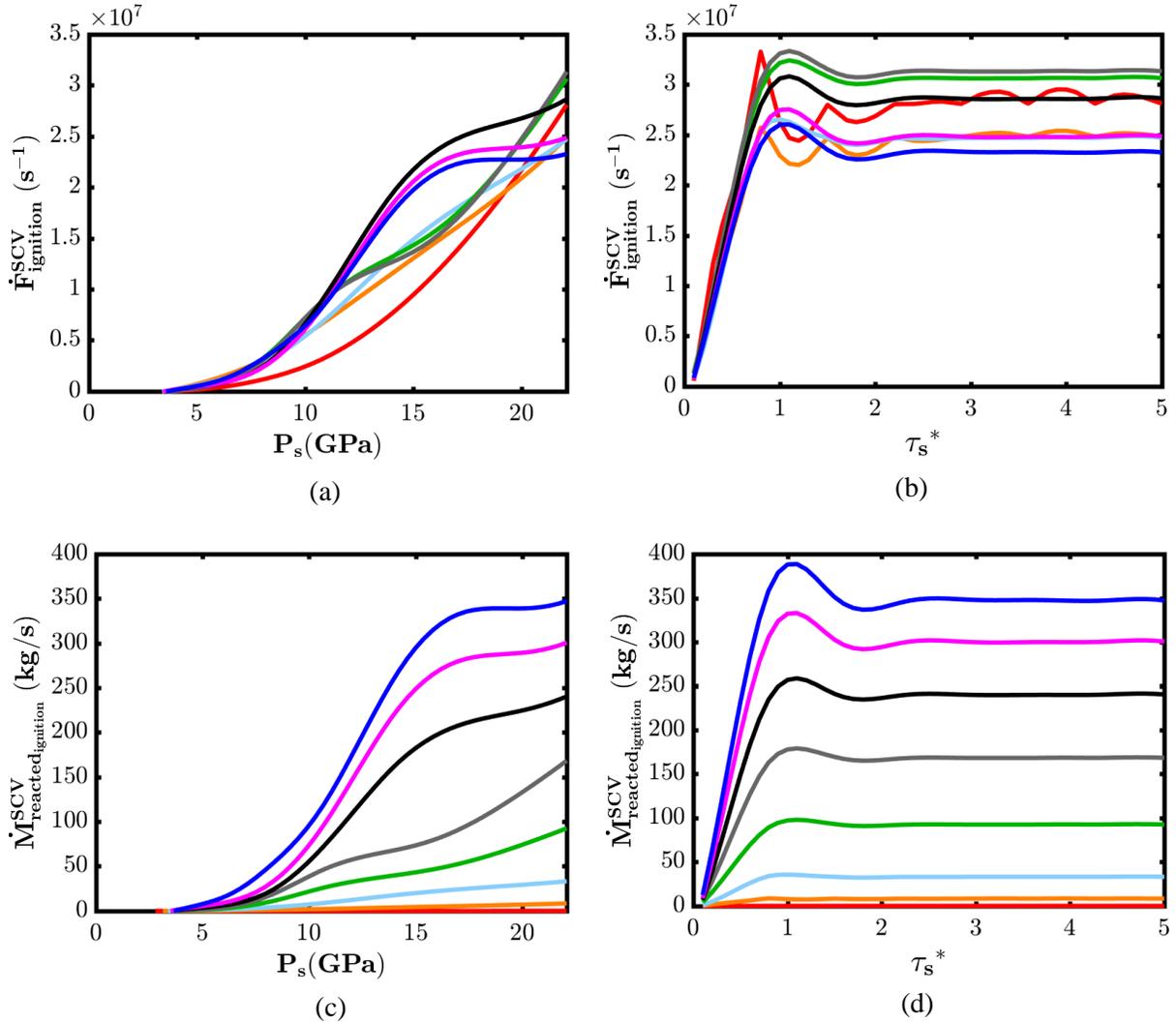

Figure 13: Variation of $\dot{F}_{\text{ignition}}^{\text{scv}}$ for void diameters ranging from 1 to 100 µm (a) with respect to (a) shock pressure $P_s$ (at constant $\tau_s^* = 5.0$) and (b) the non-dimensionalized pulse duration $\tau_s^*$ (at constant $P_s = 22.1$ GPa). The variation of the non-dimensional ignition reaction rate $\dot{M}_{\text{reacted}_{\text{ignition}}}$ with respect to (c) $P_s$ (at constant $\tau_s^* = 5.0$) and (d) $\tau_s^*$ (at constant $P_s = 22.1$ GPa).

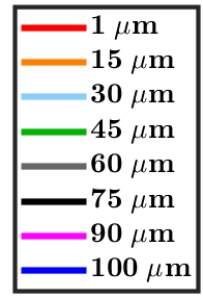

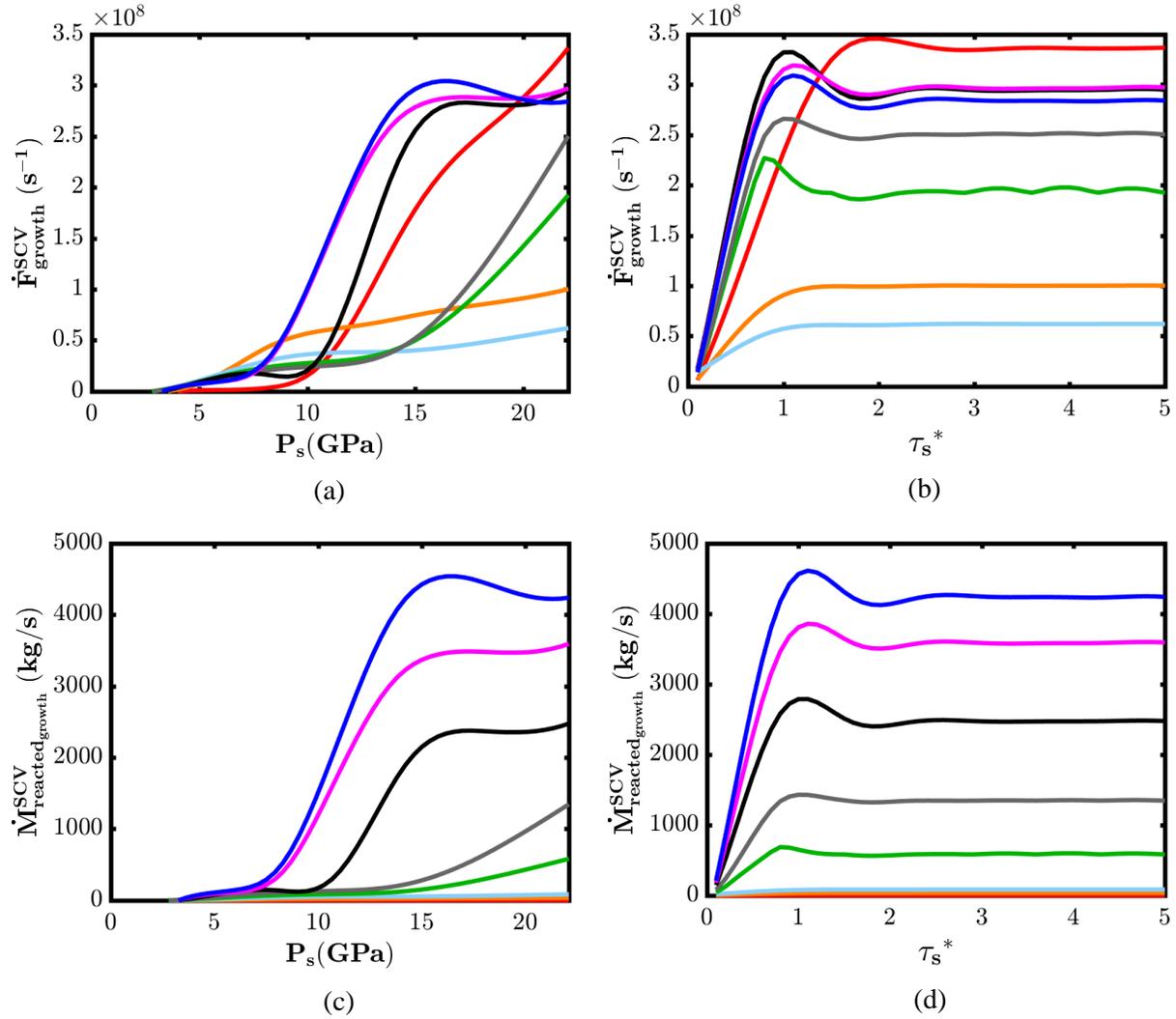

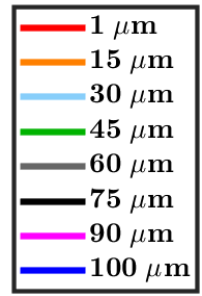

Figure 14: Variation of $\dot{F}^{scv}_{growth}$ for void diameters ranging from 1 to 100 µm (a) with respect to (a) shock pressure $P_s$ (at constant $\tau_s^* = 5.0$) and (b) the non-dimensionalized pulse duration $\tau_s^*$ (at constant $P_s = 22.1$ GPa). The variation of the non-dimensional ignition reaction rate $\dot{M}_{reacted_{growth}}$ with respect to (c) $P_s$ (at constant $\tau_s^* = 5.0$) and (d) $\tau_s^*$ (at constant $P_s = 22.1$ GPa).

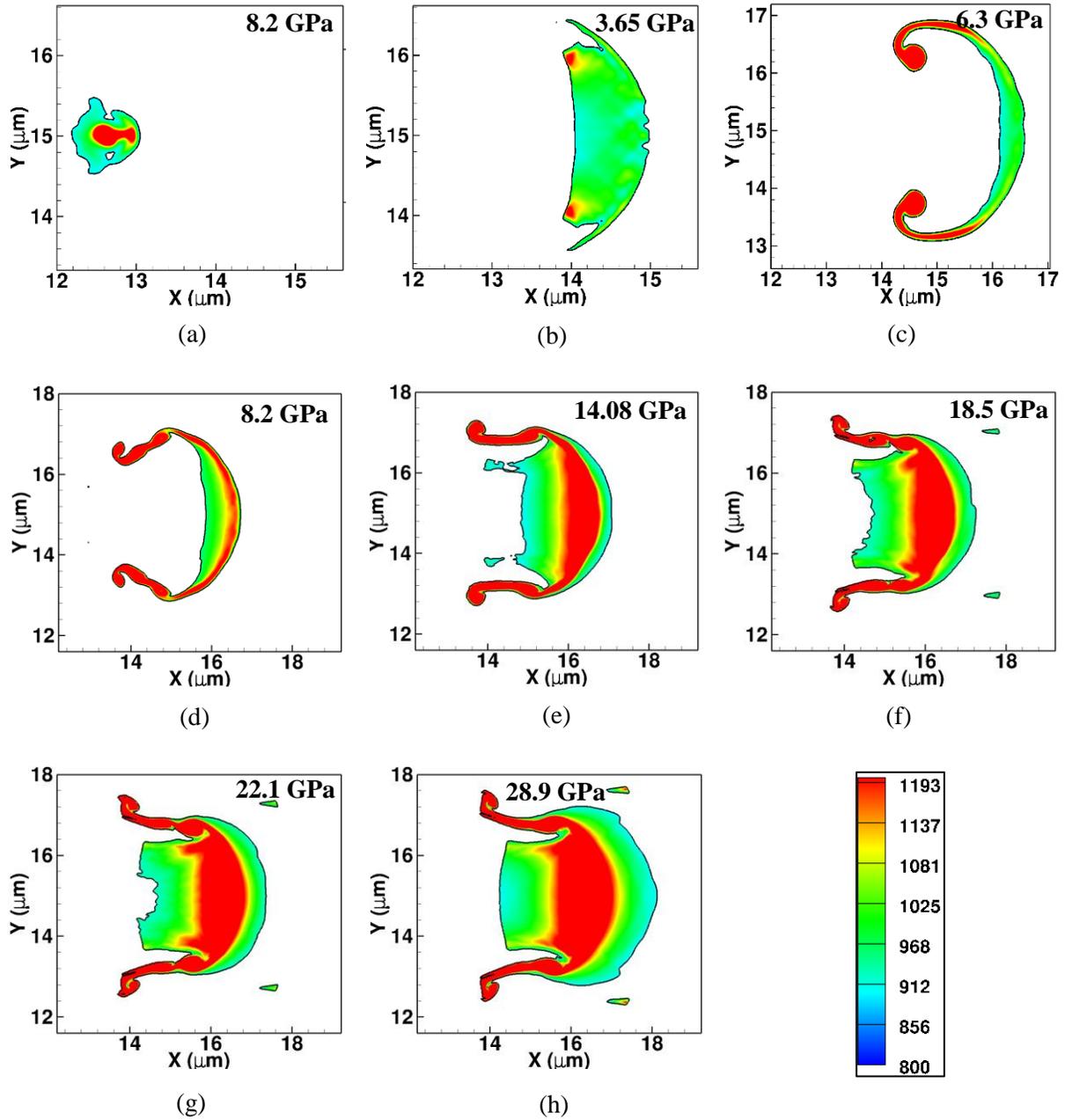

Figure 15: Contours of the hot-spot temperature (K) for a void of size, $D_{\text{void}} = 10$ μm subjected to different shock pressures $P_s$ ranging from 1.27 GPa up to 28.9 GPa in figures (a) to (h). The shock pulse duration is kept constant, and set to $\tau_s^* = 5.0$.

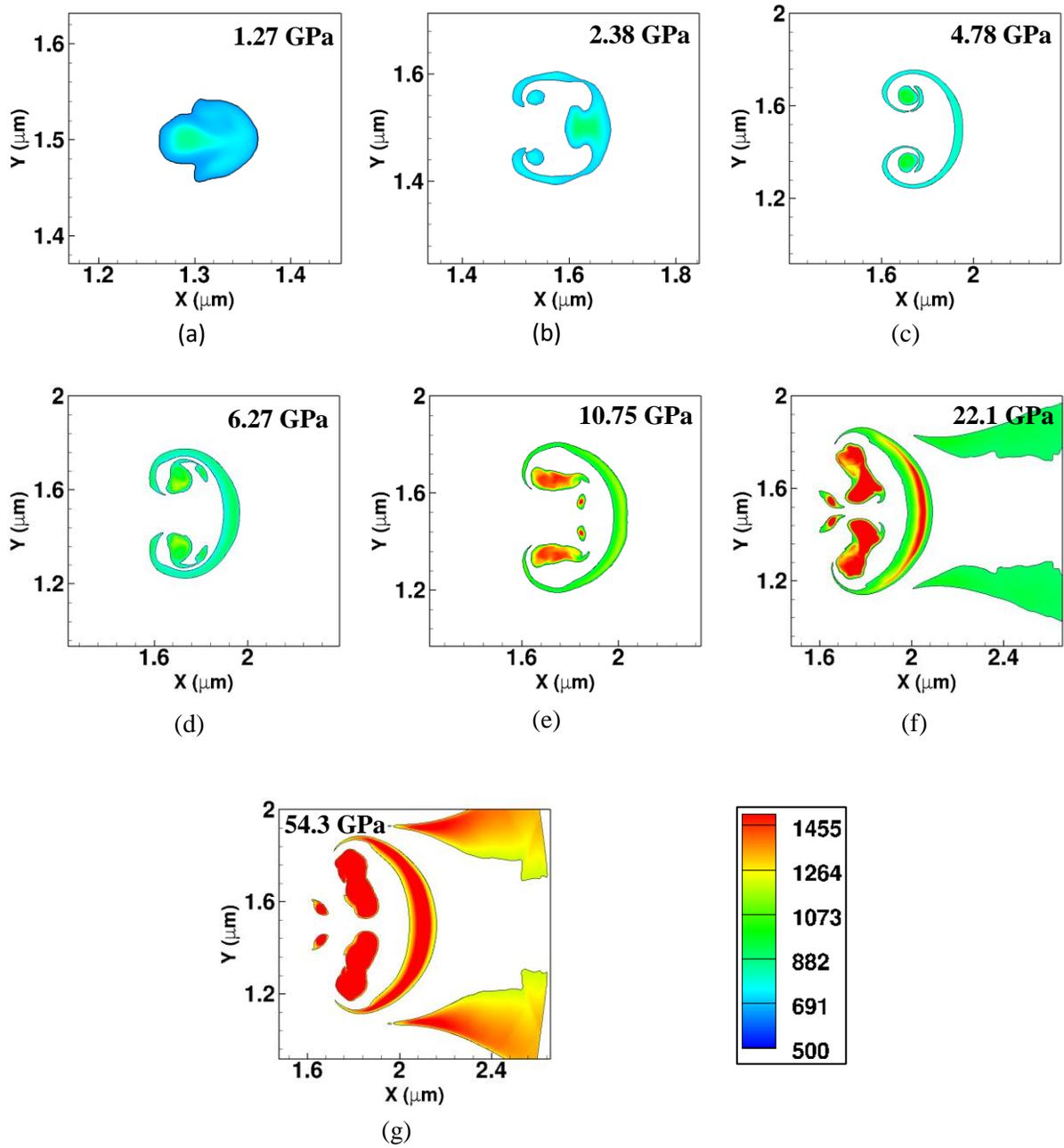

Figure 16: Contours of the hot-spot temperature (K) for a void of size, $D_{\text{void}} = 1$ µm subjected to different shock pressures $P_s$ ranging from 1.27 GPa up to 54.3 GPa in figures (a) to (h). The shock pulse duration is kept constant, and set to $\tau_s^* = 5.0$.

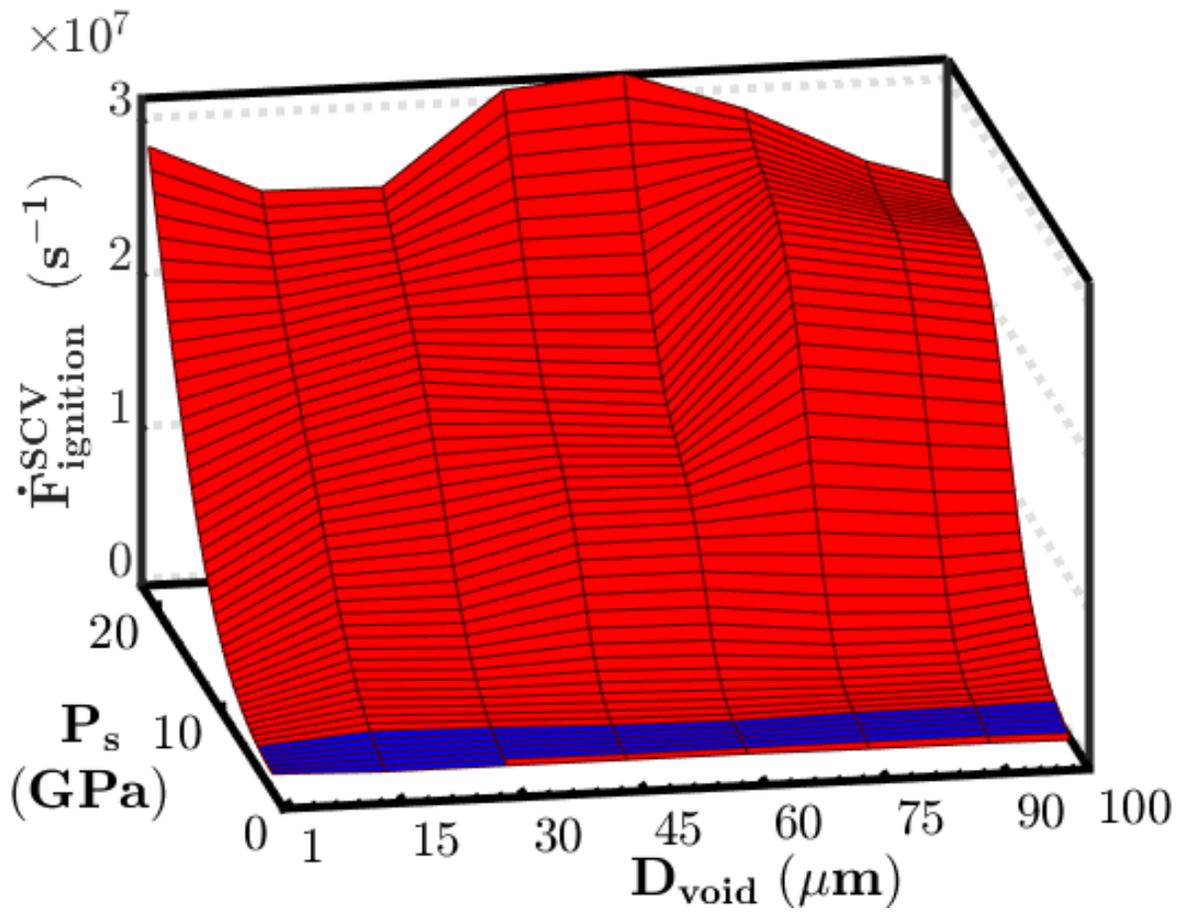

Figure 17: Variation of $\dot{F}_{\text{ignition}}^{\text{scv}}$ with respect to shock pressure $P_s$ and $D_{\text{void}}$ at $\tau_s^* = 5.0$. The blue parts of the surface indicate the sub-critical regions whereas the red ones are the critical/super-critical regions.

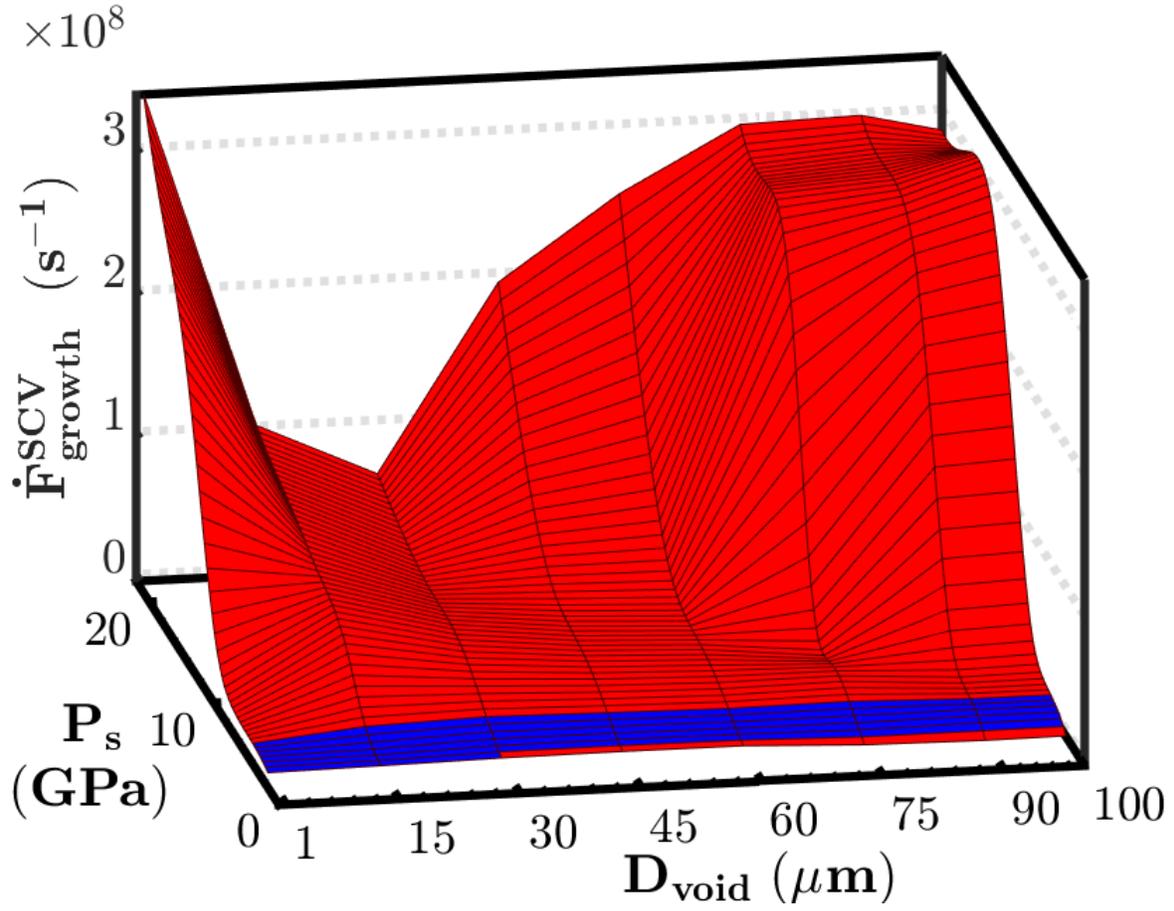

Figure 18: Variation of $\dot{F}_{\text{growth}}^{\text{scv}}$ with respect to shock pressure $P_s$ and $D_{\text{void}}$ at $\tau_s^* = 5.0$. The blue parts of the surface indicate the sub-critical regions whereas the red ones are the critical/super-critical regions.

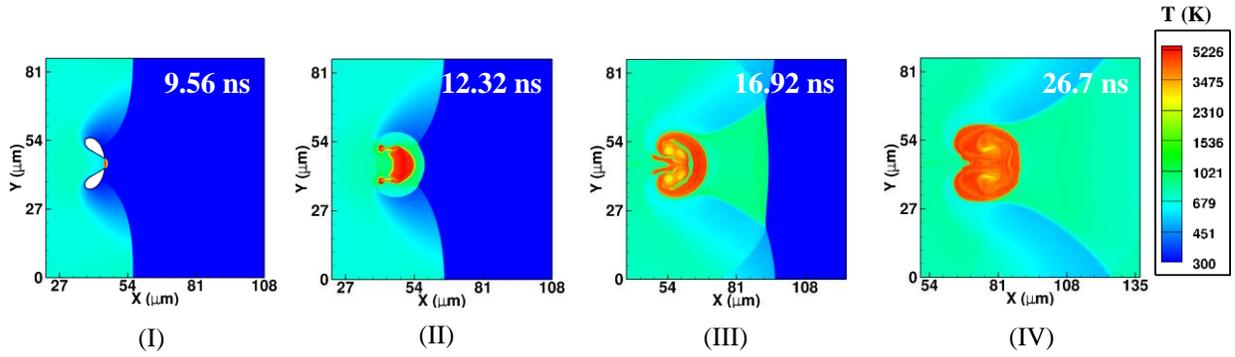

(a) Hot spot temperature contours for the collapse of a 30 μ$m$ diameter void

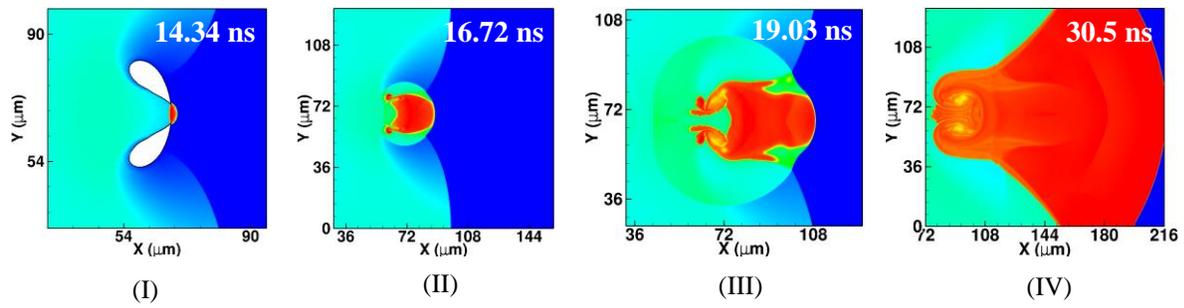

(b) Hot spot temperature contours for the collapse of a 45 μ$m$ diameter void

Figure 19: Hot spot evolution for the collapse of four diameters viz. $D_{\text{void}}$ = 30 μ$m$ and 45 μ$m$ under a sustained shock load of pressure, $P_s$ = 22.1 GPa.